\documentclass{aa}  
\usepackage{graphicx}
\usepackage{txfonts}
\usepackage{natbib,twoopt}

\bibpunct{(}{)}{;}{a}{}{,} 


\usepackage{color}
\begin{document}

   \title{Tracing planet-induced structures in circumstellar disks using molecular lines}


   \author{F. Ober\inst{1}, 
           S. Wolf\inst{1},
            A. L. Uribe\inst{2,3}
            \and
            H. H. Klahr\inst{2}
           }

   \institute{Institute of Theoretical Physics and Astrophysics, University of Kiel,
              Leibnizstraße 15, 24118 Kiel, Germany \\
              \email{fober@astrophysik.uni-kiel.de} 
              \and Max Planck Institute for Astronomy, K\"{o}nigstuhl, 69117 Heidelberg, Germany 
              \and University of Chicago, The Department of Astronomy and Astrophysik, 5640 S. Ellis Ave, IL 60637 Chicago
             }
   \date{Received, 17/03/2015 / 
         Accepted, 11/05/2015}
  \titlerunning{Tracing gaps in circumstellar disks using molecular lines}
  \authorrunning{Ober et al.}
  \abstract
   {Circumstellar disks are considered to be the birthplace of planets. Specific structures like spiral arms, gaps, and 
    cavities are characteristic indicators of planet-disk interaction. Investigating these structures can provide
    insights into the growth of protoplanets and the physical properties of the disk.} 
   {We investigate the feasibility of using molecular lines to trace planet-induced structures in circumstellar disks.}
   {Based on 3D hydrodynamic simulations of planet-disk interactions obtained with the PLUTO code, we perform 
    self-consistent temperature calculations and produce N-LTE molecular line velocity-channel maps and spectra 
    of these disks using our new N-LTE line radiative transfer code \textit{Mol3D}. 
    Subsequently, we simulate ALMA observations using the CASA simulator. 
    We consider two nearly face-on inclinations, five disk masses, seven disk radii, and two different typical 
    pre-main-sequence host stars (T~Tauri, Herbig Ae) at a distance of 140 pc. We calculate up to 141 individual 
    velocity-channel maps for five molecules/isotopoloques ($^{12}$C$^{16}$O, $^{12}$C$^{18}$O, HCO$^+$, HCN, and CS) in a total
    of 32 rotational transitions to investigate the frequency dependence of the structures indicated above.}
   {We find that the majority of protoplanetary disks in our parameter space could be detected in the 
    molecular lines considered. However, unlike the continuum case, gap detection is not straightforward in lines. 
    For example, gaps are not seen in symmetric rings but are masked by the pattern caused by the global (Keplerian) velocity field.
    By comparison with simulated observations of undisturbed disks we identify specific regions in the 
    velocity-channel maps that are characteristic of planet-induced structures.}
   {Simulations of high angular resolution molecular line observations demonstrate the 
    potential of ALMA to provide complementary information about the planet-disk interaction as 
    compared to continuum observations.
    In particular, the detection of planet-induced gaps is possible under certain conditions.}


   \maketitle
 \section{Introduction and motivation}\label{sec:intro}
 It is commonly assumed that planets form in
 circumstellar disks around young pre-main-sequence (PMS) stars \citep[e.g.,][]{Mordasini2010}.
 During this formation process planets perturb the disk because of their gravitational potential \citep{Goldreich1980}. 
 The results of this interaction are characteristic structures such as spiral arms \citep{Ward1997,Tanaka2002} 
 and gaps \citep{Papaloizou1984}. 
 Therefore, observations of these large-scale structures provide indirect constraints on the existence of 
 protoplanets and on the relevant physical processes involved in the protoplanet-disk interaction.
 \\
 \citet{Wolf2005} and \citet{Ruge2013} showed the possibility of tracing gaps in continuum re-emission light for various disk 
 configurations and ALMA configurations. In addition, \citet{Ruge2014} investigated the feasibility of tracing gaps in continuum re-emission maps and 
 scattered light images simultaneously.
 In contrast to this, only a few studies exist investigating the feasibility of tracing planet-induced structures through molecular line observations.  
 \citet{Semenov2008} showed the feasibility of drawing conclusions about disk properties such as large-scale temperature gradients and chemical stratification 
 from simulated HCO$^+$~(4-3) observations. \citet{Dutrey2008} used spatially unresolved but spectroscopically resolved line observations to 
 trace inner cavities due to the radial dependence of the (Keplerian) velocity field. 
 \citet{Cleeves2011} demonstrated the feasibility of detecting a huge inner
 cavity (R $\sim$45~AU) using molecular lines (CO, H$_2$CO$_2$). Using ALMA,
 \citet{Casassus2013a} were able to image multiple molecular gas species within the dust free inner hole of HD~142527 and to
 utilize the varying optical depths of the different species to detect a flow of gas crossing the gap opened by a planet.\\
 However, the main component of protoplanetary disks, cold molecular hydrogen (H$_2$), is very difficult to observe directly in these disks
 because of the missing electrical dipole moment. 
 Thus, other molecules with lower excitation temperatures must be observed to derive the
 dynamics, temperature, density, and chemical structure in these disks.
 The most frequently observed molecule is CO, mainly because of its high 
 abundance and low excitation energy. Observations of various isotopologues of this molecule provided the first 
 constraints on the physical properties such as the (vertical) gas temperature structure of young disks \citep[e.g.,][]{Pietu2007}.
 Through comprehensive surveys of protoplanetary disks around solar-type stars other molecules such as
 HCO$^+$, H$_2$CO, CS, CN, and HCN were detected \citep{Kastner1997,Oberg2010, Thi2004}.
 Recently, \citet{Chapillon2012a} provided evidence of HC$_3$N in the disks around GO Tau and MWC 480 using the IRAM 30 m telescope. 
 However, because of insufficient spatial resolution and the lack of sensitivity these observations were restricted to the simplest and most
 abundant molecules.
 The Atacama Large (Sub)-Millimeter Interferometer
 (ALMA) will potentially allow the direct study of the molecular composition and distribution
 in protoplanetary disks in the submillimeter wavelength regime both at global scale as well as in the potential planet-forming region.  
 The first ALMA observations of HD~163296 show that CO radiation originates
 from a thin layer with an opening angle of about 15$^\circ$ with respect to the disk midplane \citep{Rosenfeld2013,deGregorio-Monsalvo2013}.
 More recently with the unprecedented sensitivity of ALMA, cyclic C$_3$H$_2$ (which has been proposed as a useful probe of 
 radiation penetration in disks) has been detected in the
 disk surrounding HD~163296 \citep{Qi2013}.
 \\
 The motivation to search for gaps in protoplanetary disks using molecular line observations is given not only by the aim to indirectly detect planets,
 but even more to derive complementary constraints on the disk physics from the planet-disk interaction. 
 For example, the resulting gap structure, given by its shape and depth in the gas and dust distribution, depends not 
 only on the mass of the planet and grain size, but also on the interaction of the planet with the gas and dust phase \citep{Paardekooper2004}.
 As a result, larger dust grains ($\gtrsim$ 150~$\mu$m) are expected to decouple from the gas, pile up at the edges of the gap and significantly 
 alter the gas-to-dust mass ratio \citep{Paardekooper2006}. Moreover, we emphasize that a gap in the
 density distribution of a cirumstellar disk is not necessarily an irrevocable sign of an embedded planet.
 \citet{Flock2015} showed that MRI turbulence is sufficient to generate a gap in the gas phase without the presence of a planet. 
 Recently, \citet{Bruderer2014} were able to detect CO emission inside the dust cavity of IRS 48. They concluded that a massive inner companion 
 is the main driver for the clearing of the cavity and excluded photoevaporation and grain-growth only. 
 Therefore, complementary observations of the gas and the dust phase are  crucial in order to
 improve and verify our understanding of the predominant physics; e.g., the interaction between the gas and dust component 
 in the planet forming regions of protoplanetary disks.
 As a first step into this potentially new field of studying disk physics through planet-disk interaction, we investigate the feasibility of tracing
 planet-induced gaps through spatially and spectroscopically resolved molecular line observations.
 For this purpose we developed the N-LTE line and dust continuum radiative transfer code \textit{Mol3D}, which 
 is based on the radiative transfer code MC3D \citep{Wolf1999, Wolf2003c}.\\ 
 Our studies are based on spatial density and velocity distributions resulting from 3D hydrodynamic simulations of planet-disk 
 interaction using the PLUTO code \citep{Mignone2007}. The temperature structure and corresponding molecular line maps are 
 calculated with \textit{Mol3D}. These velocity-channel maps are used as an input for the CASA (ver. 4.2) simulator \citep{Petry2012}.
 Finally, we investigate the feasibility of tracing planet-induced gaps in the simulated ALMA maps. 
 Based on these results, our goal is to prepare future ALMA observations
 of similar protoplanetary disks. In contrast to earlier studies, which were often focused on exemplary case 
 studies \citep[e.g.,][]{Cleeves2011}, we explore a large parameter space covering various disk properties such as their total mass, 
 size, and abundant molecules of typical pre-main-sequence T~Tauri and Herbig Ae stars.\newline
 This study is structured as follows.
 In Sect. \ref{sec:mol3d} we briefly present the radiative transfer code \textit{Mol3D}, 
 discussing its underlying physical assumptions and test cases.
 In Sect. \ref{sec:para_study} we describe the study of tracing gaps using molecular lines with ALMA. Finally, 
 in Sects. \ref{sec:results} and \ref{sec:conclusion} we discuss our results and provide conclusions 
 about the feasibility of tracing planet-induced gaps though molecular line observations with ALMA.
  \section{Molecular line radiative transfer code}\label{sec:mol3d}
  In this section, we briefly describe our line and dust continuum radiative transfer code \textit{Mol3D}. 
  \subsection{Mol3D features}\label{sec:sequence_mol3d}
  The N-LTE 3D parallelized line radiative transfer code \textit{Mol3D} features full 3D spherical, cylindrical, and Cartesian grids with 
  several types of boundary spacing (e.g., logarithmic, linear).
  \textit{Mol3D} typically uses three steps to produce velocity-channel maps, spectra, dust continuum maps, and spectral energy distributions (SED).
  In the first step, the dust temperature is calculated based on the assumption of local thermal equilibrium. 
  To achieve a high efficiency, we use a combination of the
  continuous absorption method proposed by \citet{Lucy1999} and immediate temperature correction by \citet{Bjorkman2001}.\\
  In the second step the level populations are calculated. As discussed in Appendix 
  \ref{sec:levelpop}, this is the main problem of line radiative transfer. In the case of circumstellar
  disks, the problem can be significantly simplified by using adequate approximation methods. 
  \textit{Mol3D} allows  the local thermodynamic equilibrium (\textbf{LTE}), the free-escape probability (\textbf{FEP}),
  or the large velocity gradient (\textbf{LVG}) method to be applied. 
  It has been shown that LTE is a reliable assumption for lower transitions of abundant molecules,
  but it is recommended that LVG be used, especially for high transitions of complex molecules with low abundances \citep{Pavlyuchenkov2007}.\\
  The last step is to generate velocity spectra or velocity-channel maps using a
  new efficient ray-tracing algorithm. Here, we have to consider that the source function is not constant inside a single cell,
  which is due to the underlying velocity field. \\
  As our code solves the radiative transfer equation including
  both gas and dust, we have to take the optical properties of both into account. 
  For the intensity integration, we implemented an embedded Runge-Kutta-Fehlberg solver 
  of order 4(5) with automatic step control. 
  The main advantage is that with providing a relative ($10^{-8}$) and an absolute ($10^{-20}$) error limit for the intensity integration, the
  algorithm calculates the corresponding step-width automatically.
  Conservation of energy is ensured by a recursive pixel refinement algorithm based on the underlying model grid.\\
  \textit{Mol3D} is tested and benchmarked against other line and continuum radiative transfer codes. 
  The results are presented in Appendix \ref{sec:appendix} along with a detailed derivation of the underlying concepts and physical assumptions.
  The code will soon be publicly available\footnote{\url{https://github.com/florianober/Mol3D}} under an open-source license.
 \section{Model setup}\label{sec:para_study}
 In this section we investigate the feasibility of observing planet-induced structures using molecular lines with ALMA.  
 For this purpose we perform 3D hydrodynamic (HD) simulations, follow-up 
 line-radiative-transport (LRT) calculations, and simulate ALMA observations using the CASA simulator.
\subsection{Three-dimensional hydrodynamic simulations}\label{sec:hydromodel}
 \begin{figure*}
   \begin{minipage}{0.5\textwidth}
         \includegraphics[width=\columnwidth]{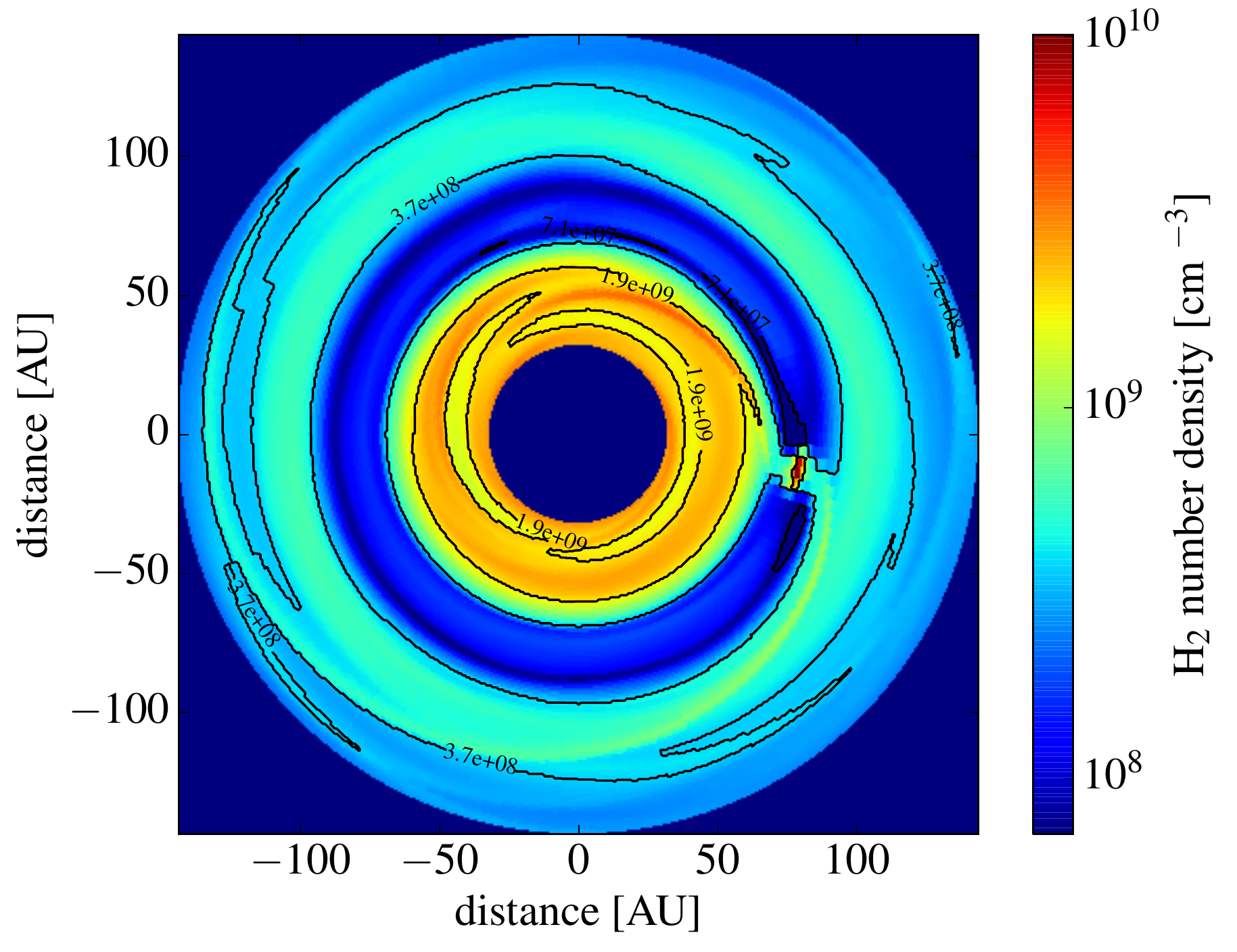}
   \end{minipage}
   \begin{minipage}{0.5\textwidth}
         \includegraphics[width=\columnwidth]{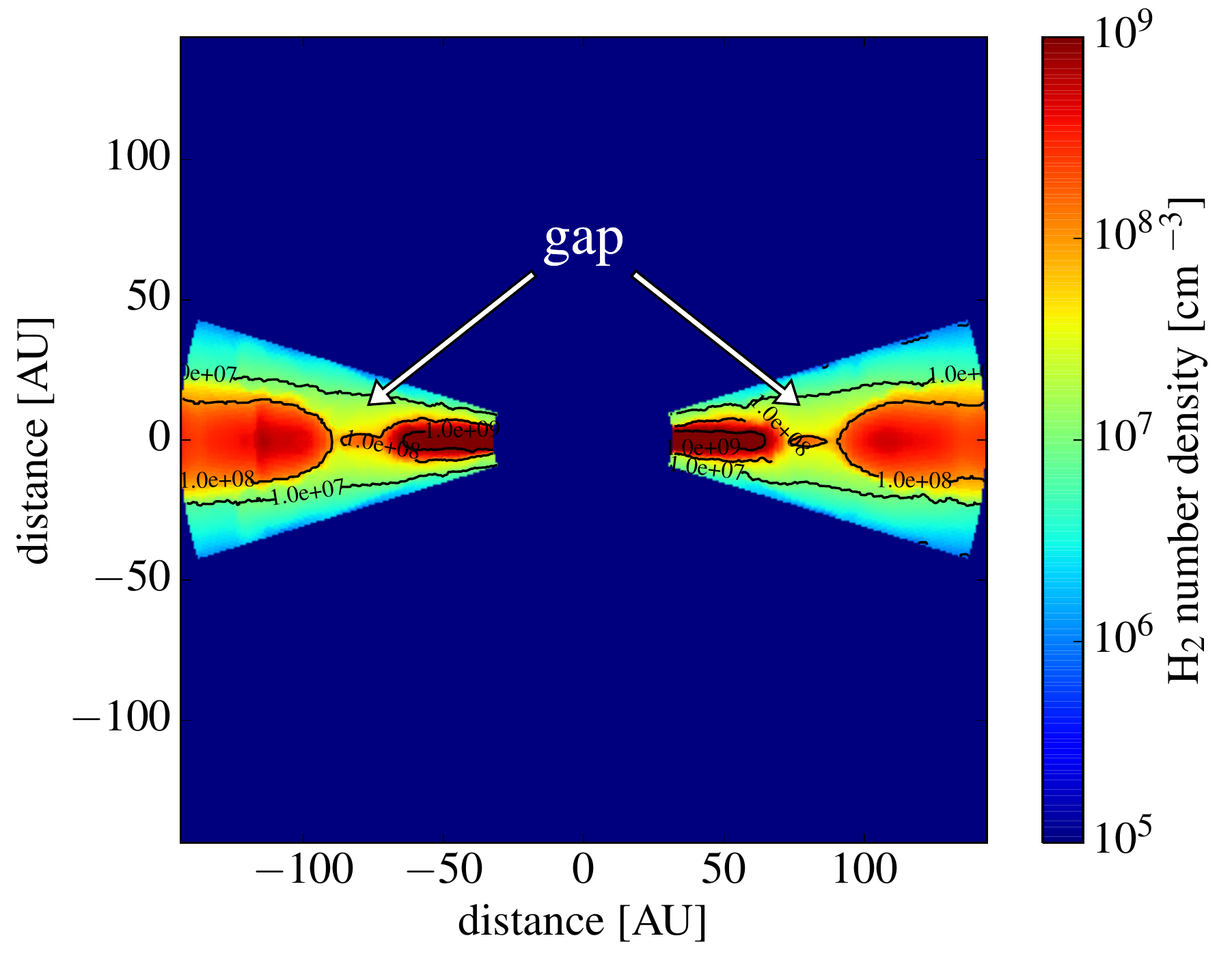}
   \end{minipage}
 \caption{Example of HD simulations with embedded Jupiter-mass  planet (planet$_{\text{mass}}$/star$_{\text{mass}}$-ratio = $10^{-3}$). 
                  Color-coded is the H$_2$ number density in log scale.
          Left: Midplane density. The planet has cleared out its orbit, producing a prominent gap.
          Owing to the perturbations, complex structures like spiral arms are also visible.    
          Right: Cut perpendicular to the disk midplane. The gap is clearly visible in the gas distribution. We note that
          the upper layers of the disk have nearly the same density as in the undisturbed case.
          }\label{pic:hd_sims}
 \end{figure*}
In preparation for this study we started with an analytic density distribution including a radially symmetric sharp gap completely depleted of dust and gas. 
We find that this approach is insufficient because it is based on an unrealistically high density contrast between the gap and its surroundings, 
resulting in a gap detection rate of almost 100\% for the planet-disk configurations considered in our study (see Sect. \ref{sec:para_lineRT}).
Instead, justified assumptions about the ratio between the density inside and outside the gap,
the gap width, and the depth of the gap are needed.\\
To reach this goal, we make use of full 3D hydrodynamic 
simulations obtained with the well-known finite-volume fluid dynamics code PLUTO \citep{Mignone2007}. 
The code uses the HLLC\footnote{Harten-Lax-van Leer-Contact \citep[see][]{Toro2009}} solver for the hydrodynamics, 
a second-order Runge-Kutta time integration and second-order variation diminishing (TVD) scheme.
The setup of the simulations was originally presented in \citet{Uribe2011} and the models used in this work are shown in \citet{Ruge2013}. 
From their entire set of planet-disk configurations we consider two scenarios. 
The first model includes an embedded Jupiter-mass planet (fixed mass ratio $M_{\text{planet}}/M_{\text{star}}$ = 0.001) on a fixed circular orbit.
The second disk with the same initial parameter setting does not contain a planet, 
and thus serves as an undisturbed smooth reference disk for comparison.\\ 
We use a spherical grid where $\theta$ is the angle from the vertical axis and $\phi$ is the azimuthal angle.
To save computation time, the vertical extent of our PLUTO simulations is chosen to cover only
$\theta$ $\in$ [$\pi$/2 - 0.3, $\pi$/2 +0.3], covering four pressure scale heights above the disk midplane.\\
As initial conditions, it is assumed that the disk gas has a sub-Keplerian rotation profile, and that the density profile $\rho_0 \propto r^{-a}$ and
the speed of sound $c_{s} = c_0(r \sin\theta)^{-b}$ follow a power-law with exponents a = 3/2 and b = 0.5, respectively.
The initial density distribution is given by
\begin{align}
\rho(r,\theta) = (r\sin\theta)^{-\frac{3}{2}}\cdot \exp\left(\frac{\sin\theta-1}{c^2_0}\right).
\end{align}
The disk is described by a locally isothermal equation of state
$P = c_s^2\rho$. The ratio of the pressure scale height $h$ to the
radial coordinate of the disk is taken to be a constant such
that $h = H /(r \sin \theta ) = 0.07$.
A softened gravitational potential is used for the planet,
\begin{align}
\Phi(r,\theta) = -\frac{GM}{\sqrt{|\vec{r}-\vec{r_p}|^2 + \epsilon^2 }}, 
\end{align}
where $\epsilon$ is the softening parameter which is defined as a fraction of the Hill radius $\epsilon = l\cdot r_p(M_p/3)^{1/3}$ with $l$ = 0.3.\\
The original disk configuration has an inner and outer edge of 2 AU and 9 AU, respectively. The planet is located at $r_p$ = 5 AU.
For the hydrodynamics simulations, buffer zones are included to avoid boundary effects, so the complete radial domain is from 1-10 AU.
For the follow-up radiative transfer, we need to convert the resulting density distribution. 
We average every two cells in each direction to save memory. 
Thus, we are able to reduce the required number of grid cells to $\frac{1}{8}$.
For the subsequent radiative transfer (RT) simulations with \textit{Mol3D} 
we add two cells (one above and one below the disk) to fill the entire $\theta$ $\in$ [0, $\pi$] domain.
The grid parameters for both codes are summarized in Table \ref{tab:code_cells}.
\begin{table}
\begin{center}
\begin{tabular}{llllll}
\hline
\hline
 code & $\theta$ [rad] & $\phi$ [rad] & N$_{r}$ & N$_{\theta}$ & N$_{\phi}$ \\
\hline
 PLUTO   & [$\pi$/2-0.3,$\pi$/2+0.3] & [0,2$\pi$] & 256 & 128 & 256 \\
 Mol3D   & [0,$\pi$]    & [0,2$\pi$] & 100 & 66 & 128  \\
\hline
\\
\end{tabular}
\caption{Comparison of grid parameter space of both codes. For the original (unscaled) model the
         radial coordinate ranges from 2 to 9~AU for the RT code and from 1 to 10~AU for the PLUTO code because of necessary buffer zones. 
         The hydrodynamic simulation results have been reduced in order to be usable with \textit{Mol3D}.
         }\label{tab:code_cells}
\end{center}
\end{table}
\subsection{Disk model}\label{sec:para_lineRT}
The hydrodynamic simulations discussed in Sect. \ref{sec:hydromodel} rely on arbitrary units for length and mass. This allows us 
to introduce linear scaling parameters to scale the disk mass and size to our needs for protoplanetary and/or transitional disks. 
\subsubsection*{Disk size}
For the inner and outer disk radius we chose the linear scaling parameter $k \in [7,..,25]$. Hence, the inner radius of the disk
is in the range of R$_{\text{in}} = k\cdot 2$~AU (i.e., R$_{\text{in}}$ $\in$ [14,...,50~AU]) 
and the outer radius in the range of R$_{\text{out}} = k\cdot 9$~AU (i.e., $\in$ [63,...,225~AU]. 
As a result, our disk models mimic different evolutionary states from early protoplanetary disks to more evolved transitional disks
or circumbinary disks with huge inner cavities.
The disk density distribution resulting from the simulation of a Jupiter-mass planet located at 80~AU in a disk
ranging from R$_{\text{in}}$ = 32~AU to R$_{\text{out}}$ = 144~AU (corresponding to a scaling parameter of k = 16) is presented in Fig. \ref{pic:hd_sims}. 
Several specific features in the disk density distribution that result from planet-disk interaction can be seen. 
The most prominent one is the distinct gap between 68 and 92 AU. In the cut through the density distribution perpendicular to the disk midplane,
the gap appears as a tunnel in the protoplanetary disk, as the planet has hardly any influence on the disk surface 
(see Fig. \ref{pic:hd_sims}, right).
Furthermore, the midplane density distribution (Fig. \ref{pic:hd_sims}, left) shows pronounced spiral arms;   these
features are barely visible in the surface of the disk in this case as well.
\subsubsection*{Mass range}
We consider a wide range of disk masses (gas+dust) from 2.67~$\cdot 10^{-1}$~M$_{\odot}$ to 2.67~$\cdot 10^{-5}$~M$_{\odot}$ which approximately
covers the range of protoplanetary and transitional disk masses derived from submm/mm observations \citep[e.g.,][]{Andrews2010, Andrews2011}.
\subsubsection*{Dust properties and stellar parameters}
For the dust we use a mixture of 62.5\% astronomical silicate and 37.5\% graphite \citep{Weingartner2001}. 
Furthermore, we assume the 
dust grains to be spherical following a grain size distribution given in Eq. \ref{eq:grainsizedist} with an exponent of d = 3.5 \citep{Dohnanyi1969}:
\begin{align}
    dn(a) \sim a^{-d} da \label{eq:grainsizedist}.
\end{align}
We consider dust
grains with a minimum radius of a$_{\text{min}}$ = 5~nm and a maximum radius of a$_{\text{max}}$ = 250~nm as found in the interstellar
medium \citep{Mathis1977}. The radiation source is the central star. 
For this study we consider a T~Tauri and a Herbig~Ae star with parameters given in Table \ref{tab:stars}).\\
\begin{table}[h]
\begin{center}
\begin{tabular}{l|l|l}
\hline
\hline
 parameter & T~Tauri & Herbig Ae \\
\hline
T$_{\text{star}}$ [K] & 4000 & 9500\\
R$_{\text{star}}$ [R$_{\odot}$] & 2 & 2.48\\
M$_{\text{star}}$ [M$_{\odot}$] & 0.7 & 2.5\\
L$_{\text{star}}$ [L$_{\odot}$] & 0.92 & 43 \\
\hline
\end{tabular}
\caption{Considered pre-main-sequence stars and their properties.}\label{tab:stars}
\end{center}
\end{table}
\subsubsection*{Distribution of the different molecular species}
For the gas-phase, we assume a gas-to-dust-mass ratio of 100:1 and perfect thermal coupling between gas and dust. \\
For the spatial distribution of different molecular species relative to each other only weak constraints are available from
observational and numerical studies.
For example,
observations have shown carbon monoxide emission from the cold outer parts of 
protoplanetary disks where CO should be frozen out \citep[$<\sim$ 20 K;][]{Dartois2003,Goto2012}. 
A possible explanation has been proposed by \citet{Aikawa2006} and \citet{Semenov2006}. 
They argue, that the CO abundance could be significantly higher in the cold
(midplane) regions of protoplanetary disks owing to vertical and radial mixing. Motivated by this idea, \citet{Hersant2009} computed 
the chemical evolution of circumstellar disks.
They included grain surface reactions with and without turbulent mixing and CO photodesorption. 
However, the resulting CO column densities are still not consistent with observations.
As indicated by \citet{Cleeves2011}, the disk midplane in the gap region can be irradiated by the stellar radiation and X-rays. Thus, CO would not be frozen out
onto grains, which would result in better gap detection rate thanks to higher contrast.\\
Because the debate on relative molecular abundances in circumstellar disks is still open, we decided to use fixed abundances relative to molecular 
hydrogen.
In particular, we follow \citet{Pavlyuchenkov2007} and assume their reported abundances for all molecules considered in this study 
(see Table \ref{tab:molecules}). 
These values are obtained by using a gas-grain chemical model with surface reactions \citep{Semenov2004,Semenov2005}, 
which is based on the UMIST 95 database of gas-phase reactions \citep{Millar1997}.
This approach keeps the model simple and hence reduces the complexity of our analysis.
 \begin{table*}
 \begin{center}
 \begin{tabular}{l|c|c|c|c|c|c}
 \hline
 \hline
  molecule & transition  & frequency & ALMA  & abundance & sensitivity $\sigma$ & PWV \\
   & & [GHz] & band & N/N(H) &  [mJy] & [mm]\\
 \hline
  & (1-0) & 115.27 & 3 &  & 7.46 & 5.2\\
  & (2-1) & 230.54 & 6 &  & 3.14 & 1.3\\
 $^{12}$C$^{16}$O & (3-2) & 345.80 & 7 & 1$\cdot$10$^{-4}$ & 4.12 & 0.66\\
  & (4-3) & 461.04 & 8 &  & 11.66 & 0.47\\
  & (6-5) & 691.47 & 9 &  & 38.02 & 0.47\\
 \hline
  & (1-0) & 109.78 & 3 &  & 3.68 &  2.7\\
  & (2-1) & 219.56 & 6 &  & 3.11 &  1.3\\
 $^{12}$C$^{18}$O & (3-2) & 6.2 & 7 & 2$\cdot$10$^{-7}$ & 8.6 & 0.66\\
  & (4-3) & 439.09 & 8 &  & 61.96 & 0.47\\
  & (6-5) & 658.55 & 9 &  & 38.66 & 0.47\\
 \hline
 & (1-0) & 89.19 & 3 &  & 3.87 & 5.2\\
 & (3-2) & 267.56 & 6 &  & 3.04 & 0.91\\
 HCO$^{+}$ & (4-3) & 356.73 & 7 & 1$\cdot$10$^{-8}$ & 4.63 & 0.66\\
 & (5-4) & 445.90 & 8 &  & 1087.7 & 0.47 \\
 & (7-6) & 624.21 & 9 &  & 137.81 & 0.47 \\
 & (8-7) & 713.34 & 9 &  & 112.10 & 0.47 \\ 
 \hline
 & (1-0) & 88.63 & 3 &  & 3.89 & 5.2\\
 & (3-2) & 265.89 & 6 &  & 2.97 & 0.91 \\
 HCN & (4-3) & 354.51 & 7 & 1$\cdot$10$^{-9}$ & 4.48 & 0.66 \\
 & (5-4) & 443.12 & 8 &  & 46.78 & 0.47\\
 & (7-6) & 620.30 & 9 &  & 4.89$\cdot$10$^{5}$ & 0.47\\
 & (8-7) & 708.88 & 9 &  &  59.91& 0.47 \\ 
 \hline
 & (2-1) & 97.98 & 3 &  & 3.38 & 2.7\\
 & (3-2) & 146.97 & 4 &  & 3.17 & 1.8\\
 & (5-4) & 244.94 & 6 &  & 2.9 & 0.91\\
 & (6-5) & 293.91 & 7 &  & 3.82 & 0.91\\
 CS & (7-6) & 342.88 & 7 & 1$\cdot$10$^{-8}$ & 4.12 & 0.66\\
 & (8-7) & 391.85 & 8 &  & 9.77 & 0.47\\
 & (9-8) & 440.80 & 8 &  & 27.2 & 0.47\\
 & (10-9) & 489.75 & 8 &  & 32.11 & 0.47\\
 & (13-12) & 636.53 & 9 &  & 35.96 & 0.47\\
 & (14-13) & 685.44 & 9 &  & 33.13 & 0.47\\
 \hline
 \end{tabular}
 \caption{Properties of the different molecules considered in this study. We note that the sensitivity is calculated with the ALMA sensitivity
          calculator (version: April 2015) for an observation of three hours considering 50 antennas in \textit{dual}-polarization mode. 
          The sensitivity strongly depends on the atmospheric transmission, which explains the high values for some transitions. 
          We use the recommended values for the precipitable water vapor (PWV) column density for each transition, which are also 
          included in the sensitivity calculator.}\label{tab:molecules}
 \end{center}
\end{table*}
\subsubsection*{Line radiative transfer setup}
The level populations are calculated by using the LVG method (see Appendix \ref{sec:LVG}) which has proven its reliability for protoplanetary disks
\citep{Pavlyuchenkov2007}.
Using these results, we produce synthetic velocity-channel maps and spectra for every model configuration. We assume a pure Keplerian
rotation profile. In this study we fix the turbulent velocity $v_{\text{turb}}$ = 100~m/s for all of our models, following the 
results of earlier molecule observations \citep[e.g.,][]{Pietu2007,Chapillon2012}.
We choose a fixed bandwidth of 50~m/s and 
calculate 141 velocity-channel maps for each of the models with 10$^{\circ}$ inclination and 71 for every face-on orientated model in our sample.
An overview of the considered parameter space of our model is compiled in Table \ref{tab:paraspace}.
\begin{table}[h]
\begin{center}
\begin{tabular}{l|l}
\hline
\hline
 parameter & value(s) \\
\hline
R$_{\text{in}}$ [AU] & 14 - 50\\
R$_{\text{out}}$ [AU] & 63 - 225\\
M$_{\text{disk}}$ [M$_{\odot}$] & $2.67\cdot10^{-1, -2, ..., -5}$\\
&\\
$\rho_{\text{dust}}$ [g/cm$^3$] & 2.5\\
a$_{\text{min}}$ [$\mu$m] & 0.5 \\
a$_{\text{max}}$ [$\mu$m] & 250 \\
&\\
inclination [$^{\circ}$] & 0, 10 \\
$v_{\text{turb}}$ [m/s] & 100 \\
distance [pc] & 140 \\
observing time [h] & 3 \\
baseline [km] & 0.7 - 16.3 \\
bandwidth $\Delta$v [m/s] & 50 \\
\hline
\end{tabular}
\caption{Parameter space of the disk model.}\label{tab:paraspace}
\end{center}
\end{table}
  \section{Results}\label{sec:results}
\subsection{Ideal line maps}\label{sec:individaleffects}
In  this section we present ideal velocity-channel maps, analyze the appearance of planet-induced disk perturbations, and
investigate the feasibility of tracing these structures with ALMA (Sect. \ref{sec:ALMA_results}).
In the following we use the term ``depth of the gap'' as the flux density contrast in the velocity-channel maps caused by the gap.
\subsection*{Optical depth effects}
\begin{figure*}
\includegraphics[width=\textwidth]{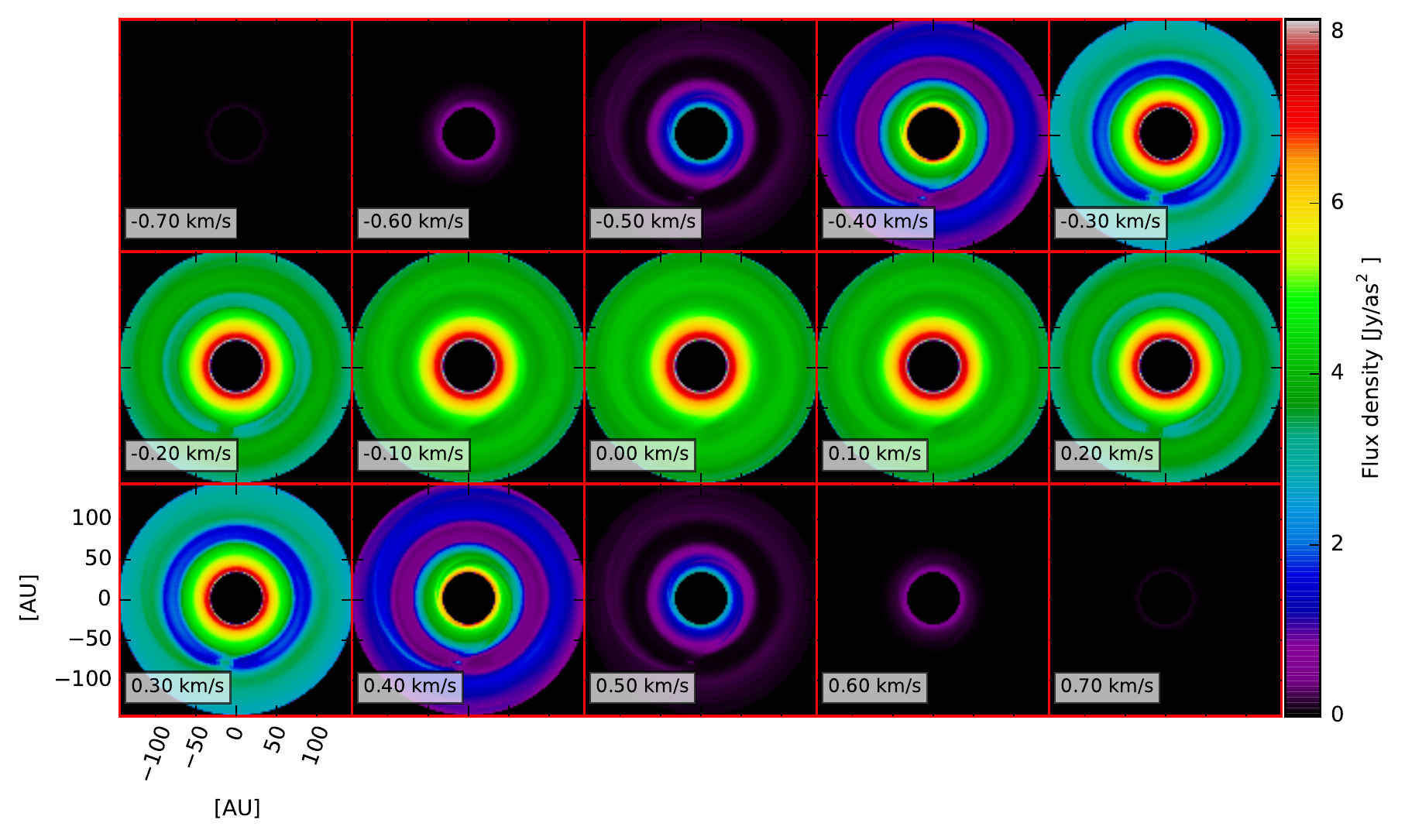}
\caption{Ideal $^{12}$C$^{16}$O~(3-2) velocity-channel map of a T~Tauri disk model with an embedded Jupiter-mass planet with an orbital distance of 80 AU.
         The disk mass amounts to $2.67 \cdot 10^{-5}$~M$_{\odot}$, which is the lowest disk mass in our sample, and the outer radius is 144 AU. 
          The flux density in Jy/mas$^2$ is color coded for each
         individual channel. The disk is seen \textit{\emph{face-on}} (0$^{\circ}$). The gap is clearly visible, particularly
         in the line wings. Even more complex 
         features like the huge spiral arm (e.g., at 300~m/s) can be recognized. The total width of the line is only given by
         thermal and microturbulent velocities (Section \ref{sec:gas_phase}).
          }\label{pic:sp1_k05m05st00i0CO3_velo_ch_map}
\end{figure*}

\begin{figure*}
\includegraphics[width=\textwidth]{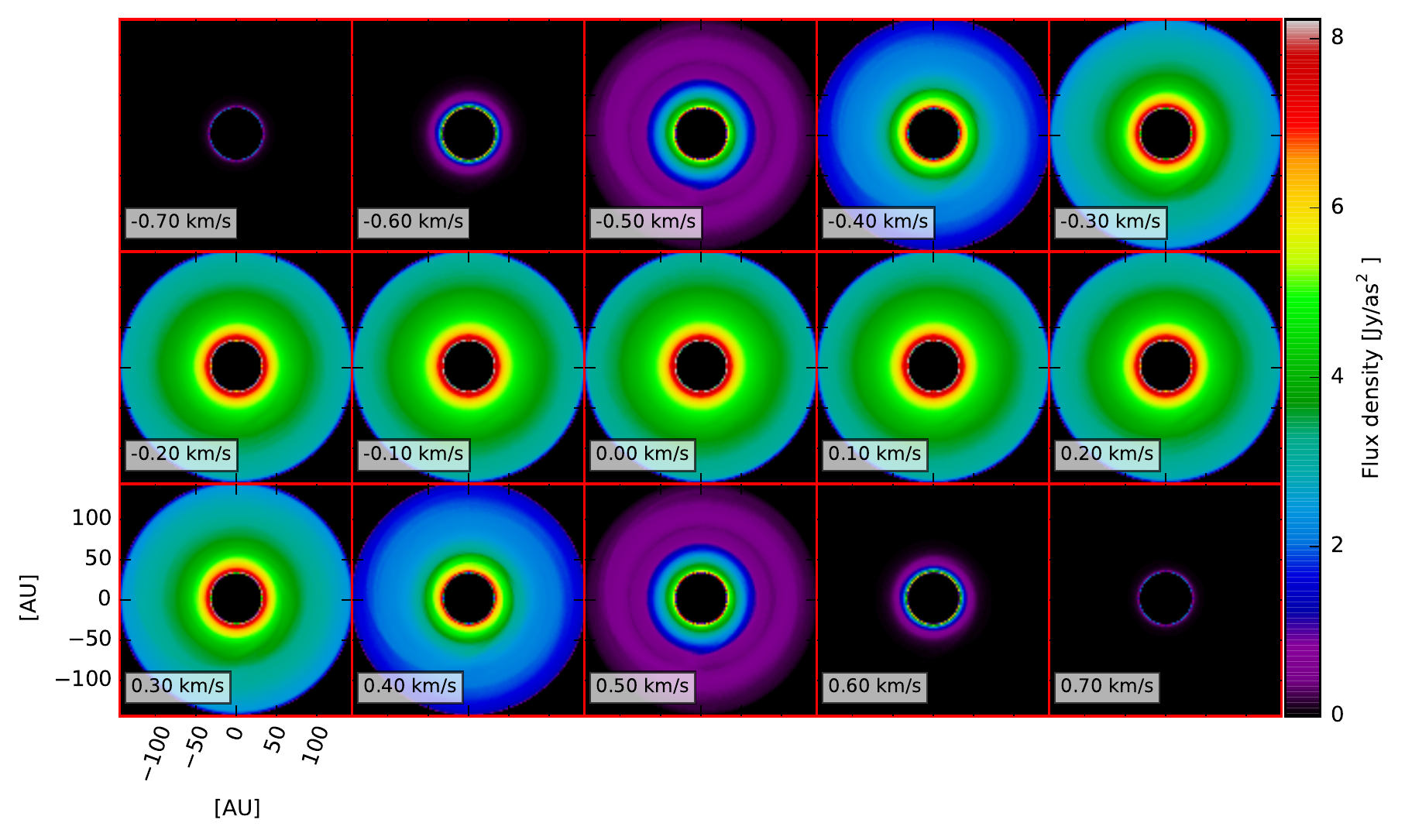}
\caption{Ideal $^{12}$C$^{16}$O~(3-2) velocity-channel map of a T~Tauri disk model with an embedded Jupiter-mass planet (at 80~AU). 
         The disk mass amounts to $2.67 \cdot 10^{-3}$~M$_{\odot}$ and the outer radius is 144~AU. 
         The flux density in Jy/as$^2$ is color coded for each
         individual channel. The disk is seen \textit{\emph{face-on}} (0$^{\circ}$). The gap is barely visible. The line is optically thick 
         owing to the material above the gap. However, the flux density is  comparable to the model with 100 times lower mass 
         (Fig. \ref{pic:sp1_k05m05st00i0CO3_velo_ch_map}).
          }\label{pic:sp1_k05m03st00i0CO3_velo_ch_map}
\end{figure*}

\begin{figure*}
\includegraphics[width=\textwidth]{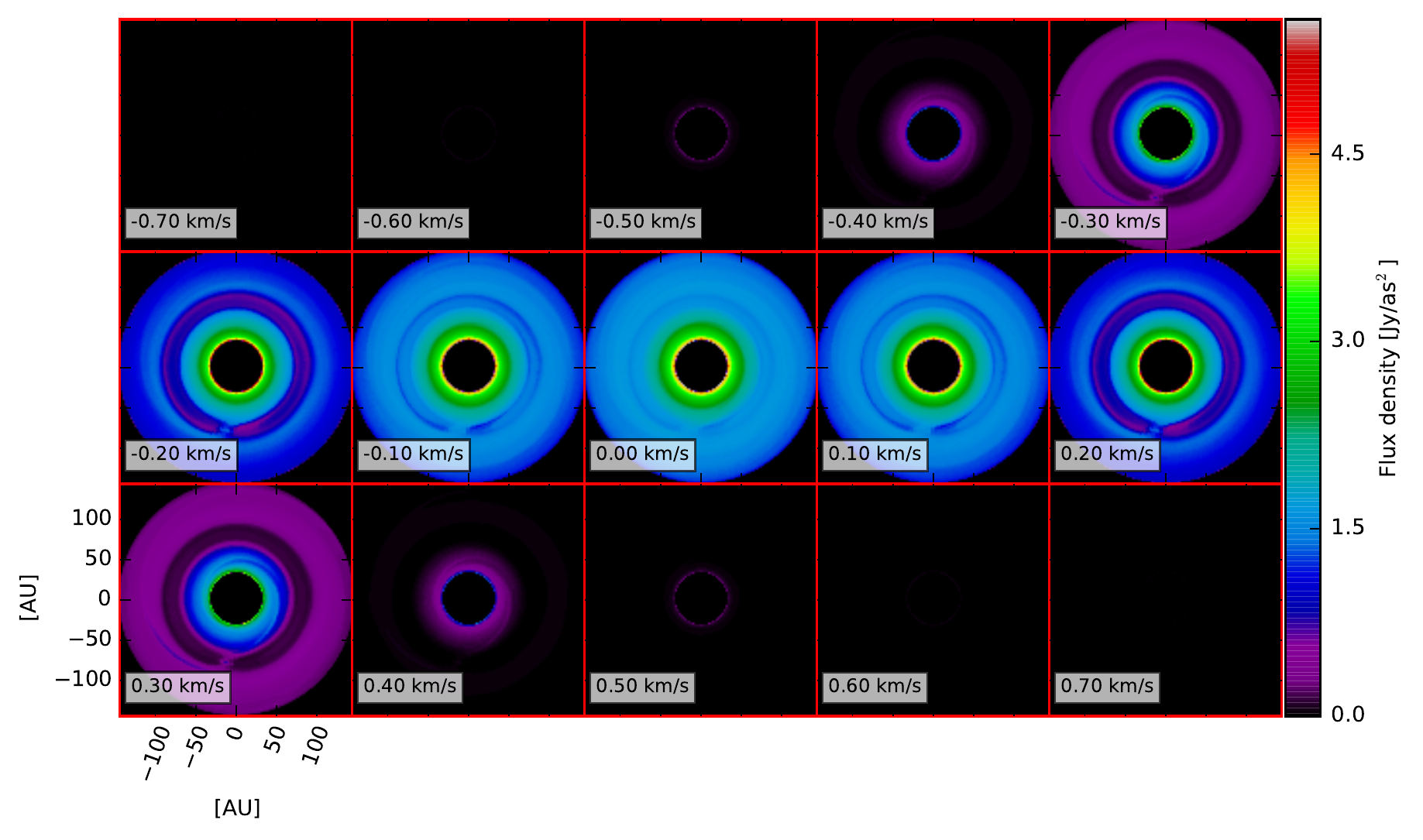}
\caption{Ideal $^{12}$C$^{18}$O (3-2) velocity-channel map of a T~Tauri disk model with an embedded Jupiter-mass planet (at 80~AU). 
         The disk mass amounts to $2.67 \cdot 10^{-3}$~M$_{\odot}$ and the outer radius is 144~AU.
         The flux density in Jy/as$^2$ is color coded for each
         individual channel. The disk is seen \textit{\emph{face-on}}. 
         Mostly because of the low abundance of this CO isotopologue, this line traces deeper parts of
         the disk. Thus, the gap is clearly detectable in contrast to the optically thick $^{12}$C$^{16}$O line of this model
         (cf. Fig.  \ref{pic:sp1_k05m05st00i0CO3_velo_ch_map}).
          }\label{pic:sp1_k05m03st00i0C18O3_velo_ch_map}
\end{figure*}

We begin our discussion based on ideal $^{12}$C$^{16}$O~(3-2) velocity-channel maps of our T~Tauri disk model. To start with, we assume a total disk mass
of $2.67 \cdot 10^{-5}$ M$_{\odot}$, which is at the very low end of typical disk masses of these 
objects \citep[e.g., HD~166191, Coka~Tau/4;][]{Kennedy2014, Forrest2004, DAlessio2005}. 
The inner rim of the chosen disk has a radius of 32~AU while the outer radius amounts to 144~AU.
The planet is located at $\sim$80~AU and the planet-induced gap has a width of $\sim$32~AU. 
The disk is seen \textit{\emph{face-on}}. In this case,
the line shows a symmetric gap throughout the different velocity channels, 
because the line width is given only by thermal and microturbulent velocities (see Fig.~\ref{pic:sp1_k05m05st00i0CO3_velo_ch_map}; see also Appendix \ref{sec:gas_phase}). 
We find that the contrast between the gap and its
edges is not constant throughout the individual velocity channels, but varies with the \textbf{optical thickness} of the chosen line. 
In the optically thin case, the gap depth  increases as it approaches the line center owing to the increasing line flux. 
In our case, the $^{12}$C$^{16}$O (3-2) transition is optically tick at its center even for this lowest disk mass of our sample.
Thus, the $\tau = 1$ surface seen in the line wings belongs to parts of the disk closer to the disk midplane than the disk layers traced 
at the line center (i.e., at 0 m/s). 
Therefore, for this specific configuration we observe the gap best at velocities $\pm$ 300~m/s, while the main contribution at the line center comes from
the upper layers of the disk. As these upper layers are less disturbed by the planet-disk interaction, 
the gap is less pronounced in the center of an optically thin line, i.e., the nearly undisturbed upper disk layers mask the gap. 
In analogy, further signatures of the planet-disk interaction like the pronounced spiral arm 
in the outer disk can be identified best in the line wings (here, in the $\pm$~400 m/s channel maps).\\
If  we increase the optical depth by increasing the disk mass by two orders of magnitude (i.e., $2.67 \cdot 10^{-3}$ M$_{\odot}$), 
which represents a typical disk mass
of protoplanetary disks, we find that the gap vanishes nearly completely in the velocity-channel maps (Fig. \ref{pic:sp1_k05m03st00i0CO3_velo_ch_map}). 
As only the envelope above the gap can be observed, the disk appears to be undisturbed without any clear signs of planet-disk interaction.
This is the main problem when observing $^{12}$C$^{16}$O. For disk masses corresponding to typical protoplanetary disks masses we find 
that the $^{12}$C$^{16}$O molecule is \textbf{not} an adequate tracer for gaps. 
In this case, other less abundant molecules/isotopologues that allow  deeper disk layers to be observed are better suited.
We illustrate this on the basis of the $^{12}$C$^{18}$O~(3-2) transition (Fig. \ref{pic:sp1_k05m03st00i0C18O3_velo_ch_map}). 
Because of the low abundance of this carbon monoxide isotopologue, this line is less optically thick and thus traces deeper disk layers. 
However, its excitation temperatures of the lower energy levels are very similar to those of the $^{12}$C$^{16}$O molecule.
In contrast to the optically thick $^{12}$C$^{16}$O line of the same model, the gap is clearly detectable.\\
The most massive disks ($2.67 \cdot 10^{-1}$ M$_{\odot}$) in our sample do \textbf{not} show any signs of a gap in any of the
considered transitions (see Table \ref{tab:molecules}).
For the reasons outlined above, for a specific molecule observation with the aim to reveal gaps,
we recommend  estimating the optical thickness of the disk (for example from its mass derived through continuum 
observations) in order to choose the appropriate molecule and (optically thin) transition.\newline
\subsubsection*{Disk inclination}
\begin{figure*}
\includegraphics[width=\textwidth]{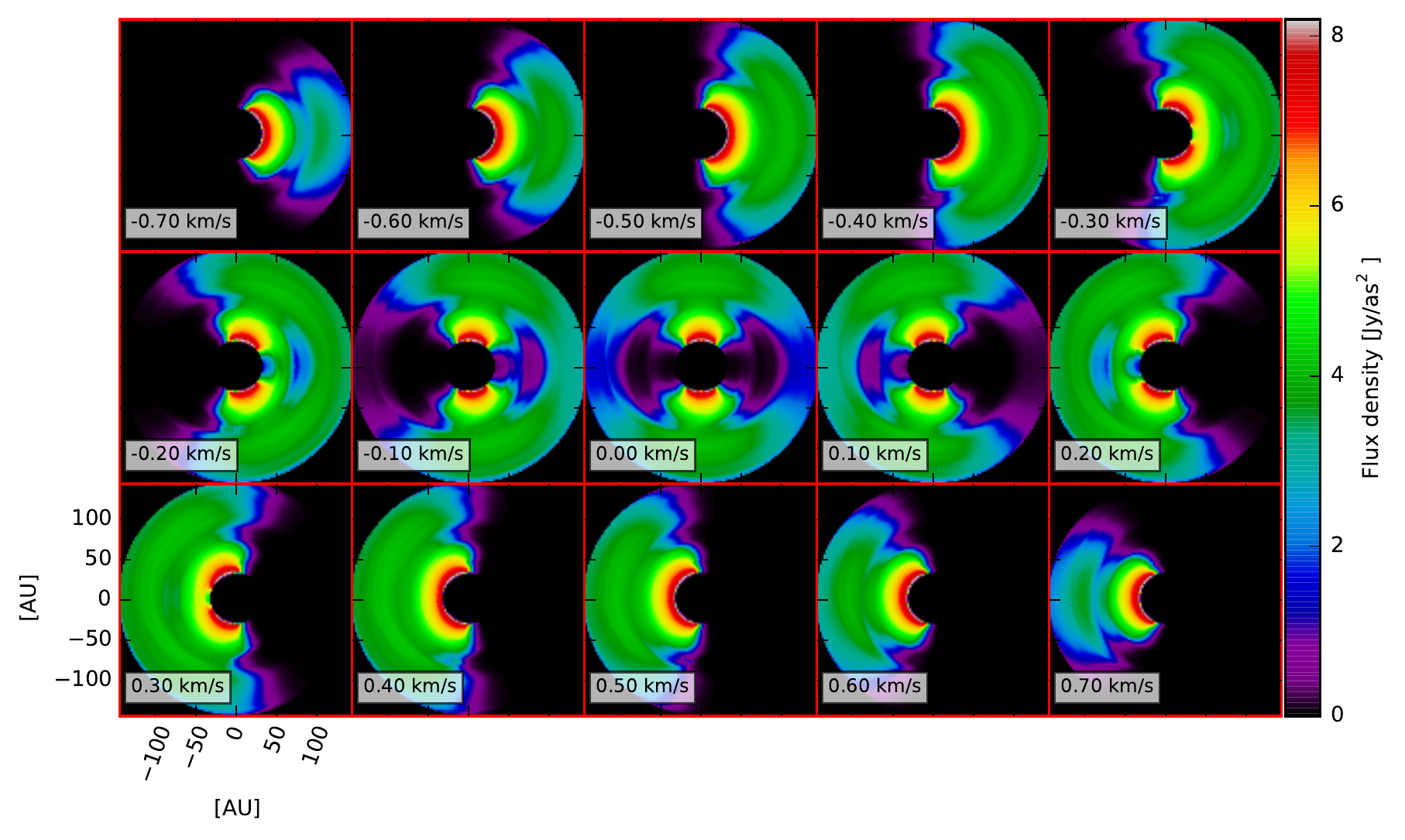}
\caption{Ideal $^{12}$C$^{12}$O (3-2) velocity-channel map of a T~Tauri disk model with an embedded Jupiter-mass planet. 
         The flux density in Jy/as$^2$ is color coded for each
         individual channel. The inclination of the disk is 10$^{\circ}$ and the disk has a total mass of $2.67 \cdot 10^{-5}$~M$_{\odot}$. 
         The gap is visible but does not appear as a symmetric ring like in the continuum 
         case because the parts of the disk contributing in the individual channel maps change due to Doppler shift.
          }\label{pic:sp1_k05m05st00i10CO3_velo_ch_map}
\end{figure*}

\begin{figure}
   \includegraphics[width=\columnwidth]{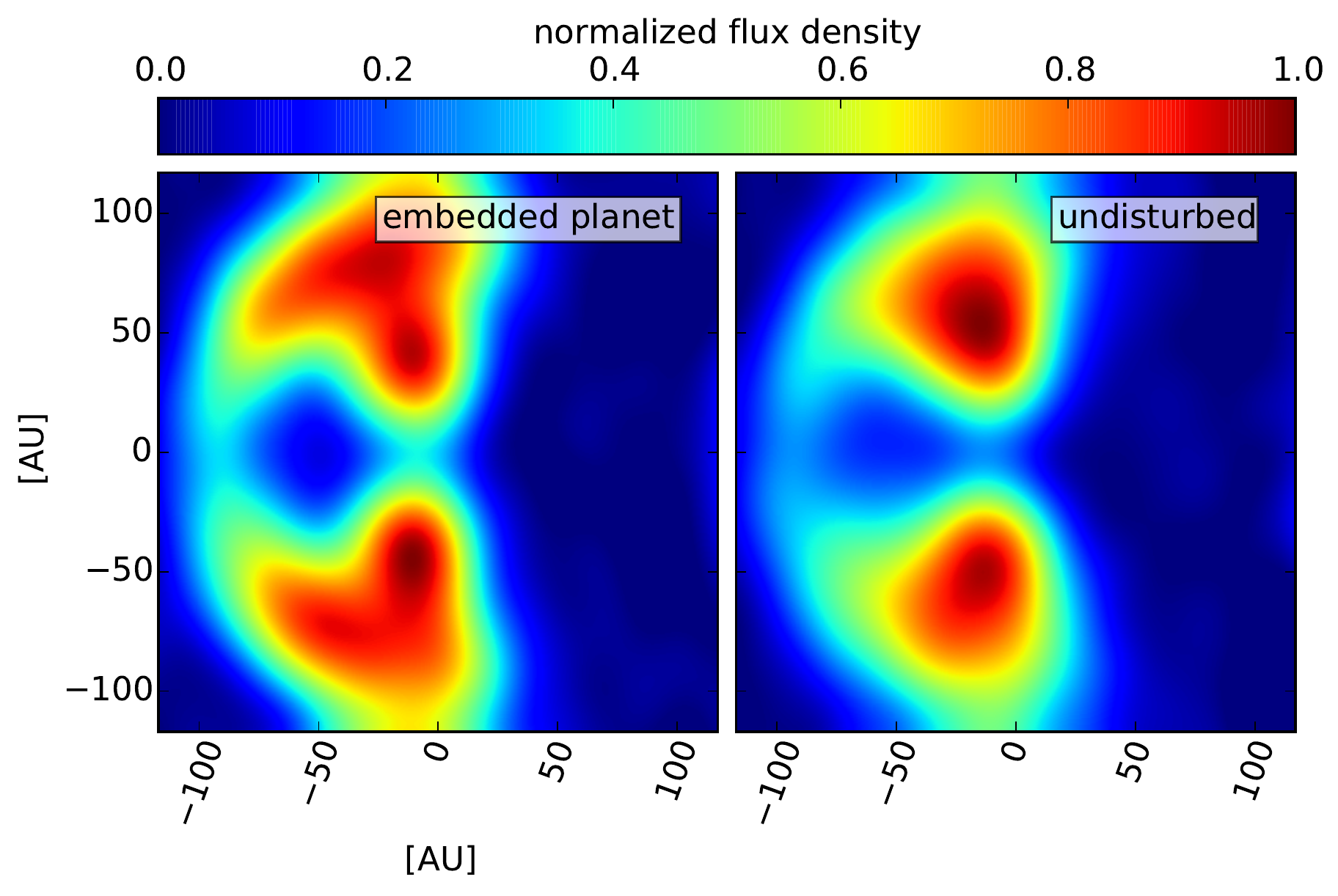}
    \caption{HCO$^+$ (4-3) velocity-channel map at 150 m/s. The disk has a slight inclination of 10$^{\circ}$.
             Disturbed (left) and smooth disks reveal similar results. Both models show indications of a gap. Thus,
             unambiguous gap detection requires detailed analysis of the origin of those structures.
             }\label{pic:gap_detect_mimicked_map.pdf}
 \end{figure}

\begin{figure}
   \includegraphics[width=\columnwidth]{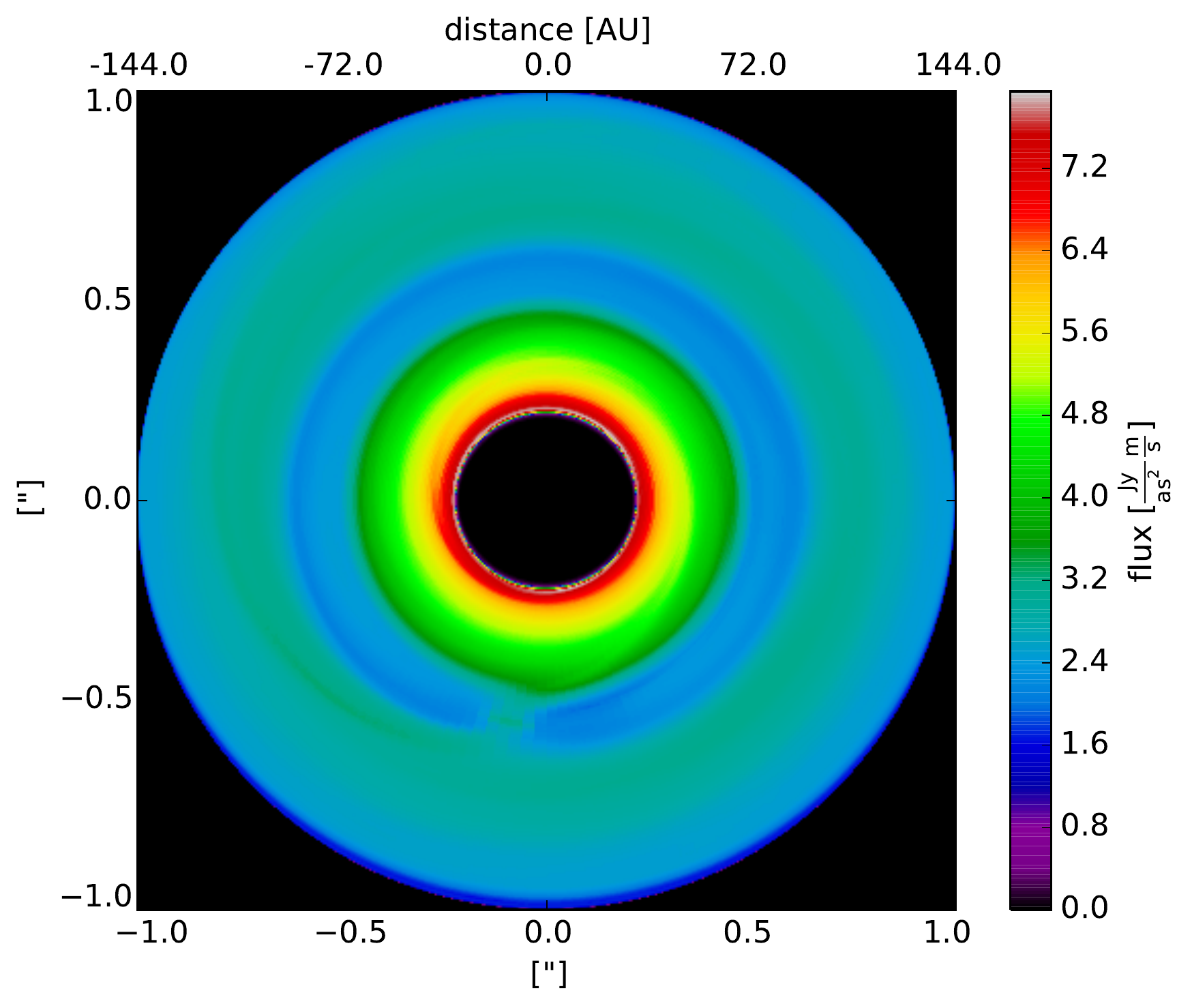}
    \caption{$^{12}$C$^{16}$O (3-2) velocity integrated intensity map of the channel map shown in Fig.~\ref{pic:sp1_k05m05st00i10CO3_velo_ch_map}.
             Owing to the integration, the patterns due to the underlying velocity field have  vanished.
             }\label{pic:velo_int_map}
 \end{figure}

In contrast to the ideal \textit{\emph{face-on}} models, the velocity-channel maps of
a disk with a small inclination of 10$^{\circ}$ is shown in Fig. \ref{pic:sp1_k05m05st00i10CO3_velo_ch_map}.
Negative velocities (top left corner) belong to parts of the disk moving toward the observer and areas with positive (bottom right)
velocities are moving away from the observer. 
In this figure the main problem for detecting gaps in inclined disks, and thus the majority of potential targets, becomes obvious.
Only fractions of the gap can be seen, but it does not appear as a symmetric ring as the gas covers a broad range of velocities
($\gg v_{\text{turb}}$) along the gap throughout the individual velocity channels. 
 This effect is illustrated in Fig.~\ref{pic:gap_detect_mimicked_map.pdf}. 
 The underlying disk model (Fig.~\ref{pic:gap_detect_mimicked_map.pdf}, left) has an outer radius of 117~AU and the planet 
 is located at 65~AU.
 Owing to the disk inclination of 10$^{\circ}$, the line is significantly Doppler shifted. The velocity-channel map shows a cavity
 in the orbital region of the planet.
 However, this not an irrevocable sign of a gap. Owing to the projected velocity in the observer's direction, 
 the undisturbed model (see Fig. \ref{pic:gap_detect_mimicked_map.pdf}, right) 
 also reveals a prominent local minimum in the same disk region. 
 Thus, the velocity patterns effectively hide the gap and hinder unambiguous identification, especially for optically thick lines.
 A possible solution is to use velocity integrated intensity maps
 for gap detection instead (see Fig.~\ref{pic:velo_int_map}). 
 For low disk inclinations, these images do not deviate much from \textit{\emph{face-on}} inclined disks, i.e., for an 
 inclination of 10$^{\circ}$ as considered in this study the flux deviation compared to the disk model seen \textit{\emph{face-on}} is less than 5\%.
 The velocity integration has the advantage that the patterns disappear and the gap can be identified easily. 
 However, the velocity dependency of the gap depth (see previous section) is lost. 
 Consequently, the velocity-integrated maps are dominated by the line flux at the line center, which traces only the upper disk layers in the case of optically
 thick lines.
\subsubsection*{Central star}
\begin{figure*}
 \includegraphics[width=\textwidth]{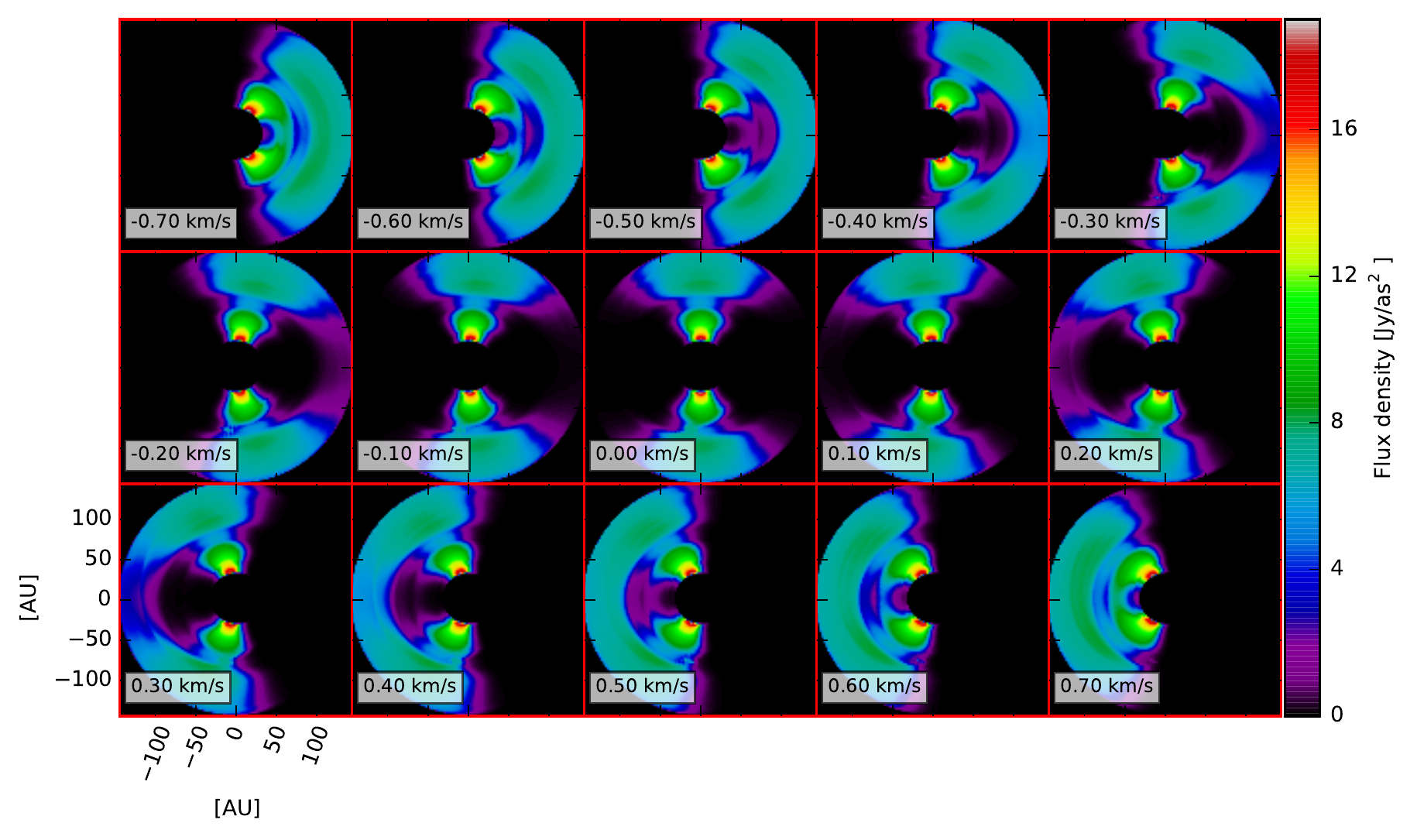}
 \caption{Ideal $^{12}$C$^{16}$O (3-2) velocity-channel map of a Herbig Ae disk model with an embedded Jupiter-mass planet.
          The disk mass amounts to $2.67 \cdot 10^{-5}$~M$_{\odot}$ and the outer radius is 144~AU.
          The flux density in Jy/as$^2$ is color coded for each
          individual channel. The inclination of the disk amounts to 10$^{\circ}$. 
          }\label{pic:sp1_k05m05st02i10CO3_velo_ch_map}
 \end{figure*}

 We now discuss the impact of using the Herbig Ae star, which is $\sim$~45 times more luminous,  as the central star.\\
 As the disk temperature is higher than
 the T~Tauri case, the molecules are exited to higher energy states. The line radiation of the outer 
 parts of the disk increases because a larger fraction of molecules is sufficiently exited to the relevant energy levels due to the warmer disk. 
 In contrast to the situation in the outer disk regions, the higher temperature in the innermost region results in a depopulation of the lower energy
 states.
 Consequently, the low line transitions we consider in this study produce less flux in this region.\\
 Second, the mass of the Herbig Ae star is about 3.5 times higher, resulting in a higher Keplerian velocity at a given radial distance from the 
 central star. Thus, for the inclined disks the linewidth of the Doppler broadened line is much higher
 than in the T~Tauri star case. Consequently, at a certain velocity channel (bandwidths of 50 m/s in this study) 
 the parts of the disk with the appropriate velocities to contribute to this channel 
 appear precise and well defined (see Fig. \ref{pic:sp1_k05m05st02i10CO3_velo_ch_map}). Thus, assuming the same disks for both PMS stars, 
 individual disk regions can be better studied in the case of the more massive central star, which increases the chances of gap detection. 
 This effect is a direct result of the dependence of the linewidth on the kinetic temperature (cf. Eq. \ref{eq:broaden}).
\subsubsection{Detectability of gaps in the ideal case}\label{sec:how_to_gap}
\begin{figure}
   \includegraphics[width=\columnwidth]{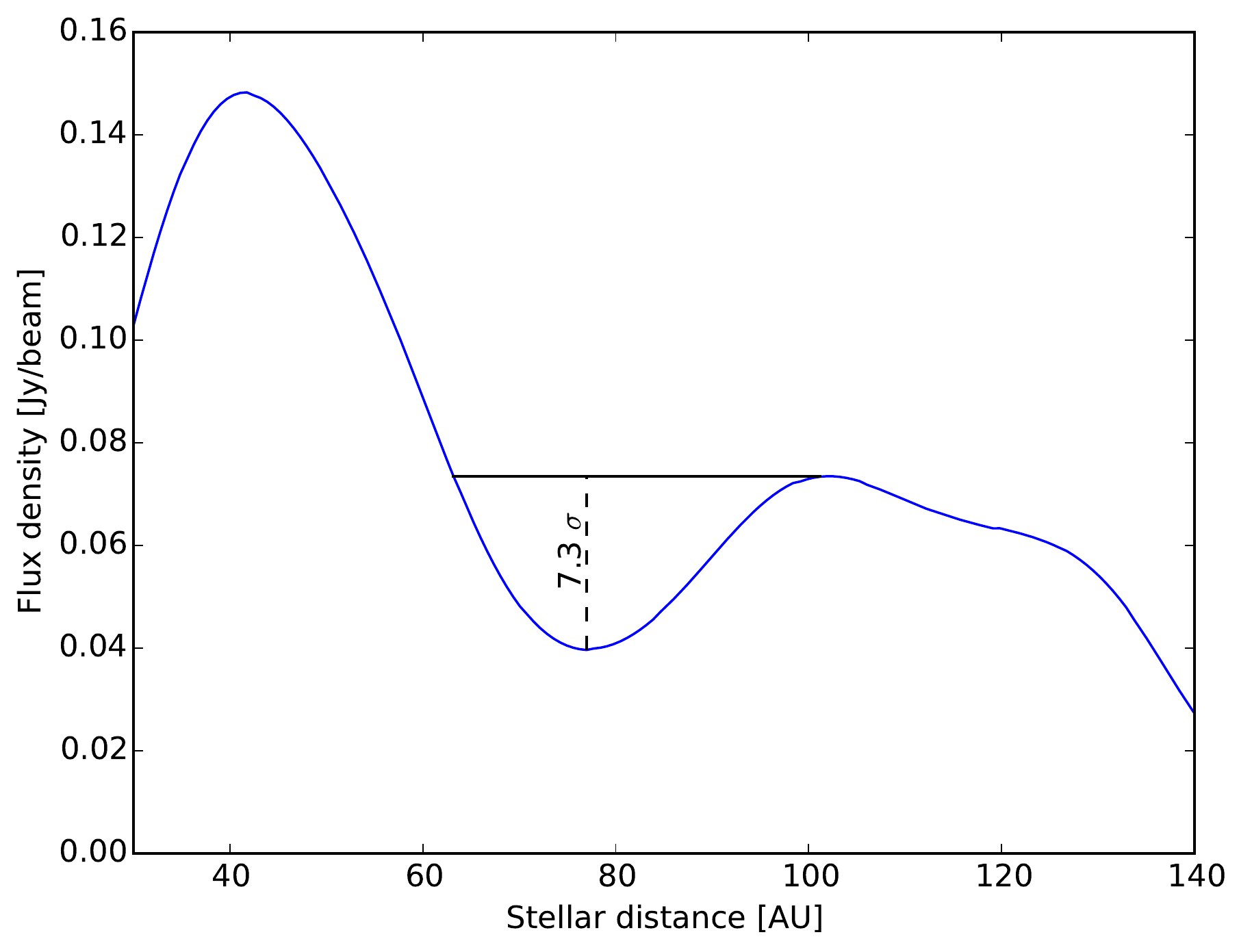}
    \caption{Illustration of the method of detecting gaps.
             The blue line indicates a radial cut through the HCO$^+$~(4-3) velocity-channel map (Fig.~\ref{pic:sp1_k05m05st00i0HCO4_velo_ch_map_alma14}) 
             at the central frequency (v = 0.0~m/s). 
             For the simulated ALMA observations, the gap is considered as being detected if the depth of the gap is deeper than three times
             the noise level $\sigma$ of the selected transition.
             In this particular case, the gap is clearly detectable, resulting in a gap contrast of $\sim7\sigma$.}\label{pic:gap_detect_image}
 \end{figure}

 We describe our approach to identifying a gap and to deriving its depth in the velocity-channel maps.\\
 We follow the procedure outlined by \citet{Ruge2013}.
 We derive the contrast between the inner/outer edge of the gap and the gap itself by using radial cuts throughout the
 simulated maps (see Fig. \ref{pic:gap_detect_image}).
 To reduce the probability of false detection (e.g., local minimum in the brightness distribution 
 due to the inclination effect discussed in Sect. \ref{sec:individaleffects}), 64 individual radial cuts are
 considered in the analysis. Only if a gap is indicated along more than 50\% of successive radial cuts is it  considered as detected.
 We successfully tested our algorithm with smooth undisturbed reference hydrodynamic models to verify that no
 substructure is misleadingly considered as a gap. 
 However, because of the very general approach due to the huge parameter space, 
 we emphasize that for specific observations better results may be achieved by adjusting  the algorithm to the specific observation
 (e.g., a well-adapted compensation of background noise/sparse uv-coverage, which leads to asymmetric structures). 
 \subsubsection*{Overview charts}
\begin{figure}
  \includegraphics[width=\columnwidth]{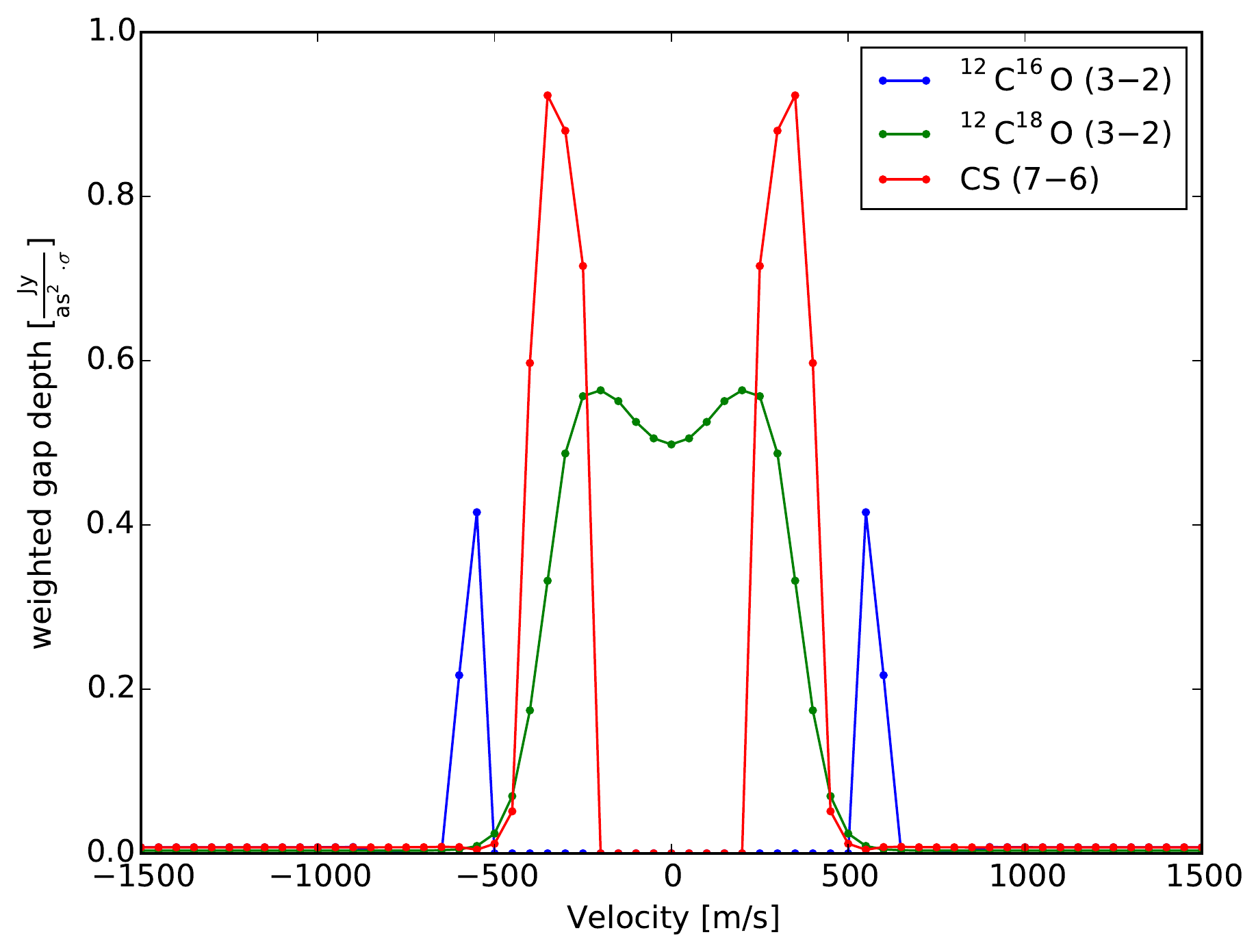}
   \caption{Depth of a gap for three different molecules/transitions over velocity and weighted by the ALMA noise level $\sigma$. 
            The disk has an outer radius of 144~AU and the disk mass 
            amounts to $2.67 \cdot 10^{-3}$~M$_{\odot}$ and the disk heated by the Herbig Ae star.
            For this disk model, all considered
            transitions are optically thick at the line center and the detection of the gap
            is only feasible with $^{12}$C$^{18}$O due to its low abundance. 
            Nevertheless, at the line wings other molecules transitions allow for gap detection too, and
            in this particular case the gap depth has a maximum value when using the CS~(7-3) transition.
            }\label{pic:k05m03st02i0_gap_best_velo}
 \end{figure}

\begin{figure*} 
   \begin{minipage}[b]{0.5\textwidth}
     \includegraphics[width=1\textwidth]{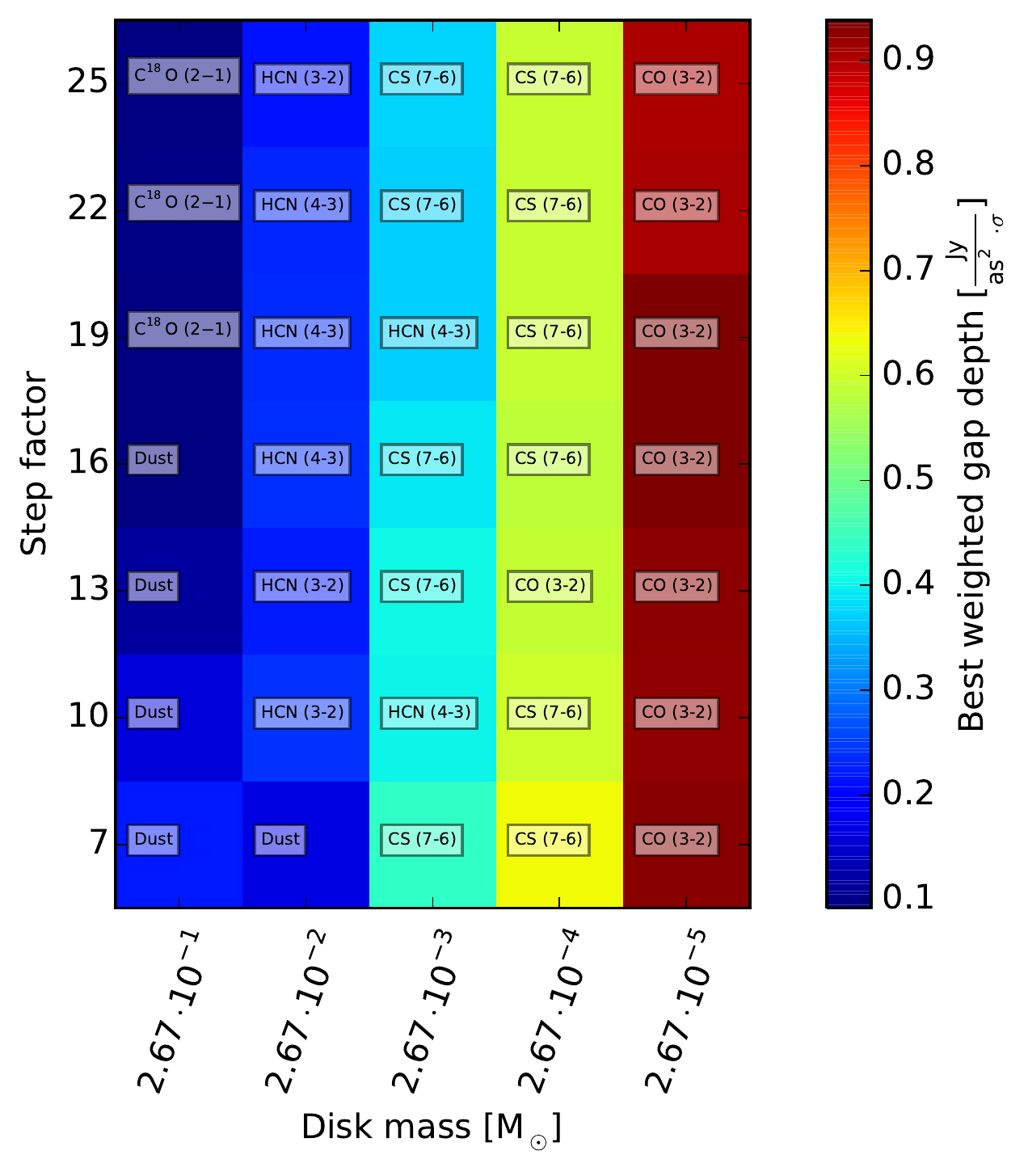}
   \end{minipage}        
   \begin{minipage}[b]{0.5\textwidth}
     \includegraphics[width=1\textwidth]{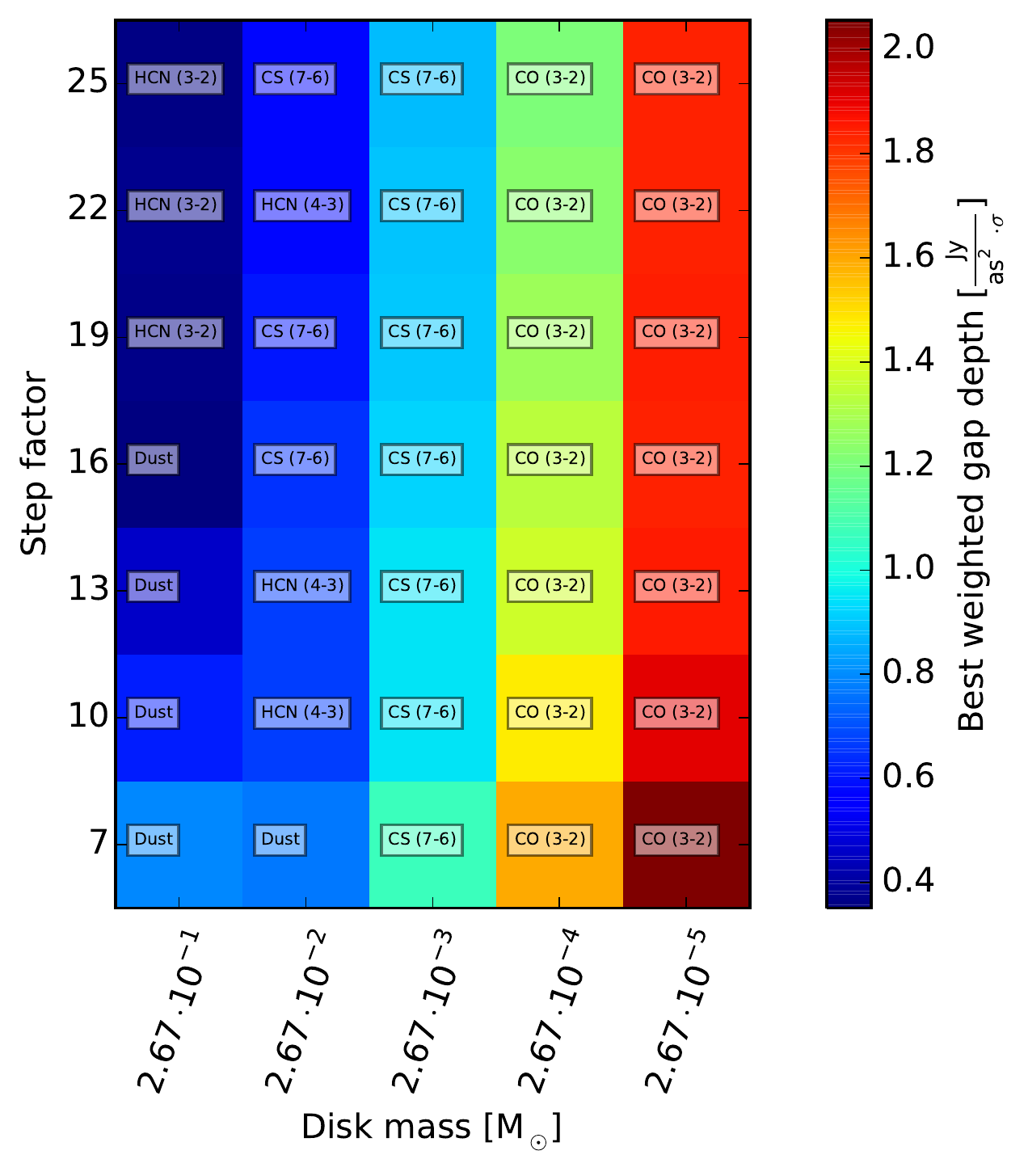}
 \end{minipage}
     \caption{Overview plot of the maximum gap depth in the ideal velocity-channel maps when comparing all transitions considered in this work. 
              The central radiation source is a T~Tauri (left) or a Herbig~Ae star (right). For the most massive disks in our sample. All considered
              transitions are optically thick and only trace the surface of the disks. Thus, for the smallest disks the gap is detectable outside the line
              in the dust continuum (marked with \textit{Dust}).
             }\label{pic:best_gap_detect_ideal}
\end{figure*}

 In  this section, the feasibility of detecting gaps through observations of selected molecular lines 
 for all considered disk configurations (see Table \ref{tab:paraspace}) is summarized.
 For this purpose, we derive the gap depth from the ideal velocity-channel maps for all five molecules and 32 assumed transitions using the algorithm described
 above. For every transition
 we derive the expected ALMA background noise of the reconstructed velocity-channel maps ($\sigma$)
 using the ALMA sensitivity calculator (see Sect.~\ref{sec:ALMA_set_up}, Table \ref{tab:molecules}). 
 Owing to the atmospheric transmission, $\sigma$ mostly rises with frequency resulting in a worse signal-to-noise ratio.
 To take care of this effect, we derive the gap depth and weight it by the expected sensitivity of ALMA at every line transition considered in this study. 
 Thus, we obtain a criterion of how well individual transitions allow gap detection with ALMA. 
 These results allow  those transitions to be chosen that are best suited for the simulated ALMA observations in Sect. \ref{sec:ALMA_results}.\\
 Figure \ref{pic:k05m03st02i0_gap_best_velo} shows the results for an exemplary disk with an outer radius of 144~AU and a disk mass of
 $2.67 \cdot 10^{-3}$~M$_{\odot}$ around the Herbig~Ae star. 
 The gap depth measured in every velocity channel for the $^{12}$C$^{16}$O~(3-2), $^{12}$C$^{18}$O~(3-2),
 and CS~(7-6) transition is shown. All transitions are optically thick at the line center. 
 While the $^{12}$C$^{16}$O~(3-2) and CS~(7-6) transitions fail to
 detect the gap at the line center, the optically thin $^{12}$C$^{18}$O~(3-2) transition allows for gap detection even at the central frequency. 
 Nevertheless, the highest gap depth is achieved by the CS~(7-6) transition at the line wings (at $\pm$350~m/s).
 \\ 
 To summarize our findings for all molecules, we search for the 
 transition that provides the highest gap depth for given disk mass and disk size and present the result in one overview map.
 The resulting charts for both considered PMSs are shown in Fig. \ref{pic:best_gap_detect_ideal}.
 The full overview charts showing the individual gap depth of all 32 transitions considered in
 this study can be found in Appendix \ref{sec:appendix_study}.\\
 As already indicated in Sect. \ref{sec:individaleffects}, 
 the most promising molecule for identifying gaps in low-mass disks ($\sim$10$^{-4} - 10^{-5}$ M$_{\odot}$)
 is $^{12}$C$^{16}$O. The gap can be detected either at the line center for the lowest disk masses or at the
 line wings for the intermediate-mass  disks. In most configurations the flux density and consequently the gap depth
 is higher even in the line wings compared to the gap depth which can be achieved with other molecules.\\ 
 For the disks with intermediate masses ($\sim$10$^{-2} - 10^{-3}$ M$_{\odot}$) of our sample gap detection becomes very challenging. 
 In these cases $^{12}$C$^{16}$O is optically thick for almost all considered disk configurations
 even at the line wings. Thus, transitions of less abundant molecules like HCO$^{+}$, HCN, or CS are better suited for gap detection. 
 However, because of the low abundance of these molecules the net flux is very low. 
 Consequently, the gap depth that relates the flux density to the background noise is too low for gap detection in many cases.
 The most massive ($\sim$10$^{-1}$ M$_{\odot}$) disks in our sample do not show any reliable sign of a gap for any molecule transition considered.
 We find that the gaps of these disks can be traced best outside the line in the dust continuum, even with the low bandwidth considered in this study.\\
 When comparing the molecular line emission from the disk around both PMSs, we find very similar results in terms of the probability of gap detection. 
 The effects due to their different stellar mass and luminosity (see Sect. \ref{sec:individaleffects} for a discussion) lead to
 slightly different molecules, which offer the best probability of detecting a gap 
 (e.g., because of different excitation states $^{12}$C$^{16}$O allows for gap detection
 even for the disks with a mass of $\sim$10$^{-4}$ M$_{\odot}$).\\
 Therefore, our results demonstrate the feasibility of gap detection using molecular lines, but it is
 only efficient for low-mass protoplanetary disks or disks with huge inner cavities (e.g., transitional disks or circumbinary disks). 
 
\subsection{Simulated ALMA observations}\label{sec:ALMA_results}
\subsubsection{ALMA simulation setup}\label{sec:ALMA_set_up}
 We use the Common Astronomy Software Application package \citep[CASA 4.2, ][]{Petry2012} to simulate observations 
 based on our ideal velocity-channel maps. 
 We consider a total of six medium and extended configurations 
 (baselines from 0.7 to 16~km) using 50~antennas, a total observing time of 3~hours, a target distance of 140~pc, 
 and the sky-coordinates of the ``Butterfly Star'' 
 \citep[$\alpha=04^{\text{h}}33^{\text{m}}17^{\text{s}}$ $\delta=+22^{\circ} 53' 20"$, ][]{Wolf2003d,Grafe2013}
 which is representative of young stellar objects in the Taurus-Auriga star-forming region. 
 The precipitable water vapor (PWV) in the atmosphere has a major influence. Therefore,
 we also include thermal noise in our simulations and use the recommended column densities for the line transitions we consider in our sample (see Table
 \ref{tab:molecules}). 
 First, we employ the \textit{simobserve} task to generate realistic visibility measurements. The final simulated observations are obtained using 
 the \textit{simanalyze} task which uses the CLEAN-algorithm for image reconstruction \citep{Hogbom1974}.
 The expected noise levels $\sigma$ are estimated with the ALMA sensitivity 
 calculator\footnote{see \url{https://almascience.eso.org/proposing/sensitivity-calculator}, version April 2015} 
 because the measured noise in the simulated images strongly depend 
 on the quality of the deconvolution\footnote{see \url{http://casaguides.nrao.edu/index.php?title=Guide_To_Simulating_ALMA_Data}}.
 The properties of the considered molecules and the expected sensitivity are summarized in Table \ref{tab:molecules}. 
 We consider the ALMA polarization configuration in \textit{dual}-mode. With respect to the results from Sect. \ref{sec:how_to_gap} 
 we select those molecules that offer the highest probability 
 of gap detection, i.e., $^{12}$C$^{16}$O$~\text{(3-2)}$, $^{12}$C$^{18}$O$~\text{(3-2)}$, CS~(7-6), HCO$^{+}~\text{(3-2)}$, and HCN~(3-2). 
 In addition, their corresponding transitions around 345 GHz (ALMA band 7, see Table \ref{tab:molecules}), offer a reliable 
 comparability of the different molecules and yield a good compromise between spatial resolution (higher frequencies) and better atmospheric 
 transmission (lower frequencies)\footnote{see \url{https://almascience.eso.org/about-alma/weather/atmosphere-model}}. 
 The $^{12}$C$^{16}$O~(4-3) transition is added to provide a second $^{12}$C$^{16}$O transition offering higher spatial resolution at the cost of 
 sensitivity.

 \subsubsection{Simulated line observations}
\begin{figure*}
   \includegraphics[width=\textwidth]{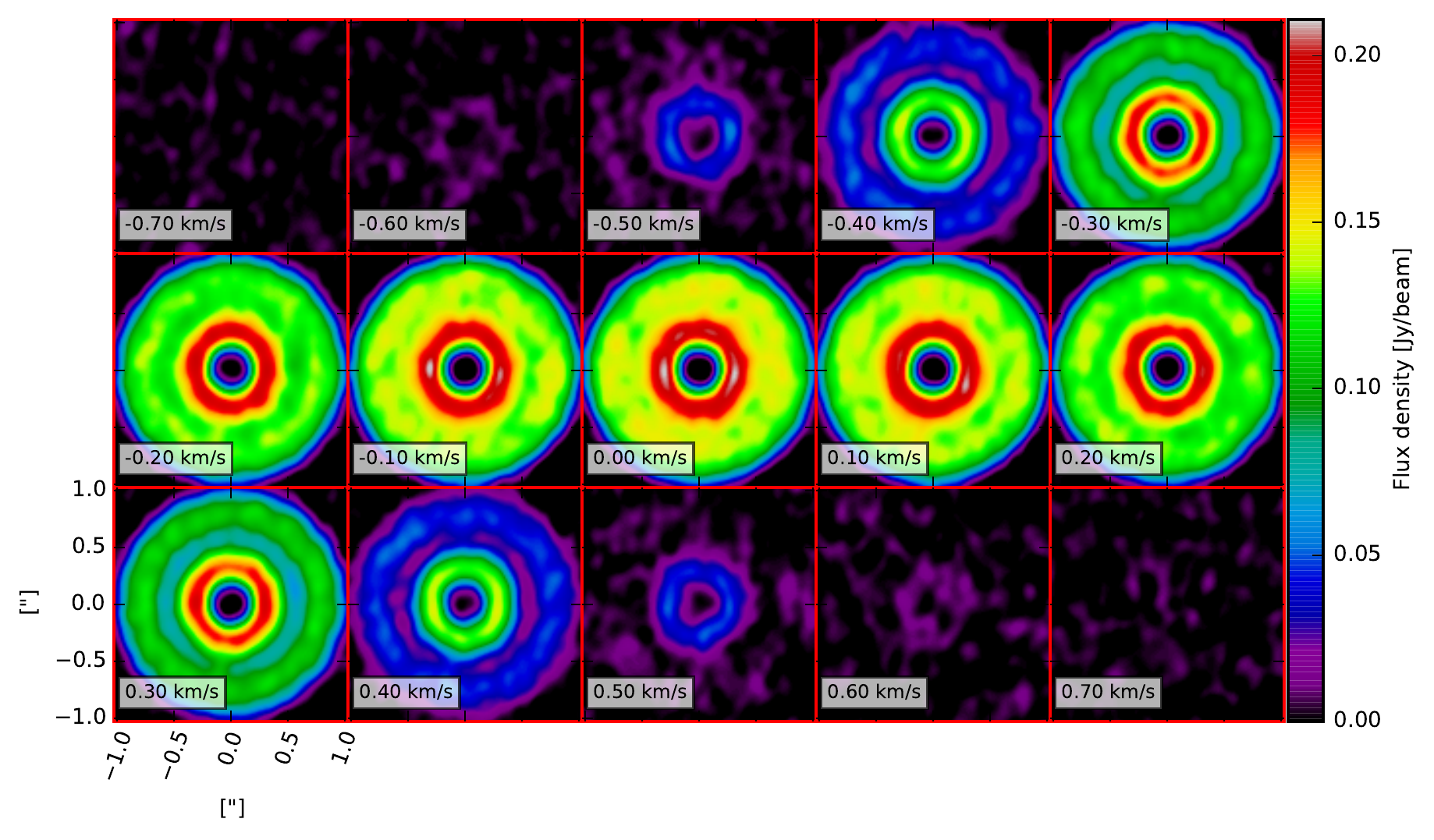}
   \caption{$^{12}$C$^{16}$O~(3-2) velocity-channel map of a model with an embedded Jupiter-mass planet as it could be 
            observed using ALMA configuration 14 (1.6 km baseline).
                        The inclination of the disk is 0$^{\circ}$ (\textit{face-on}). The outer disk radius amounts to 144~AU and 
                        the disk mass to $2.67 \cdot 10^{-5}$~M$_{\odot}$.
            The flux density in Jy/beam is color coded for each individual channel.
            In contrast to the ideal velocity-channel maps, complex structures
            are eliminated owing to beam convolution and noise. However, the gap can still be distinguished at the line slopes ($\pm$ 300 m/s).
            }\label{pic:sp1_k05m05st00i0CO3_velo_ch_map_alma14}
 \end{figure*}

\begin{figure*}
   \includegraphics[width=\textwidth]{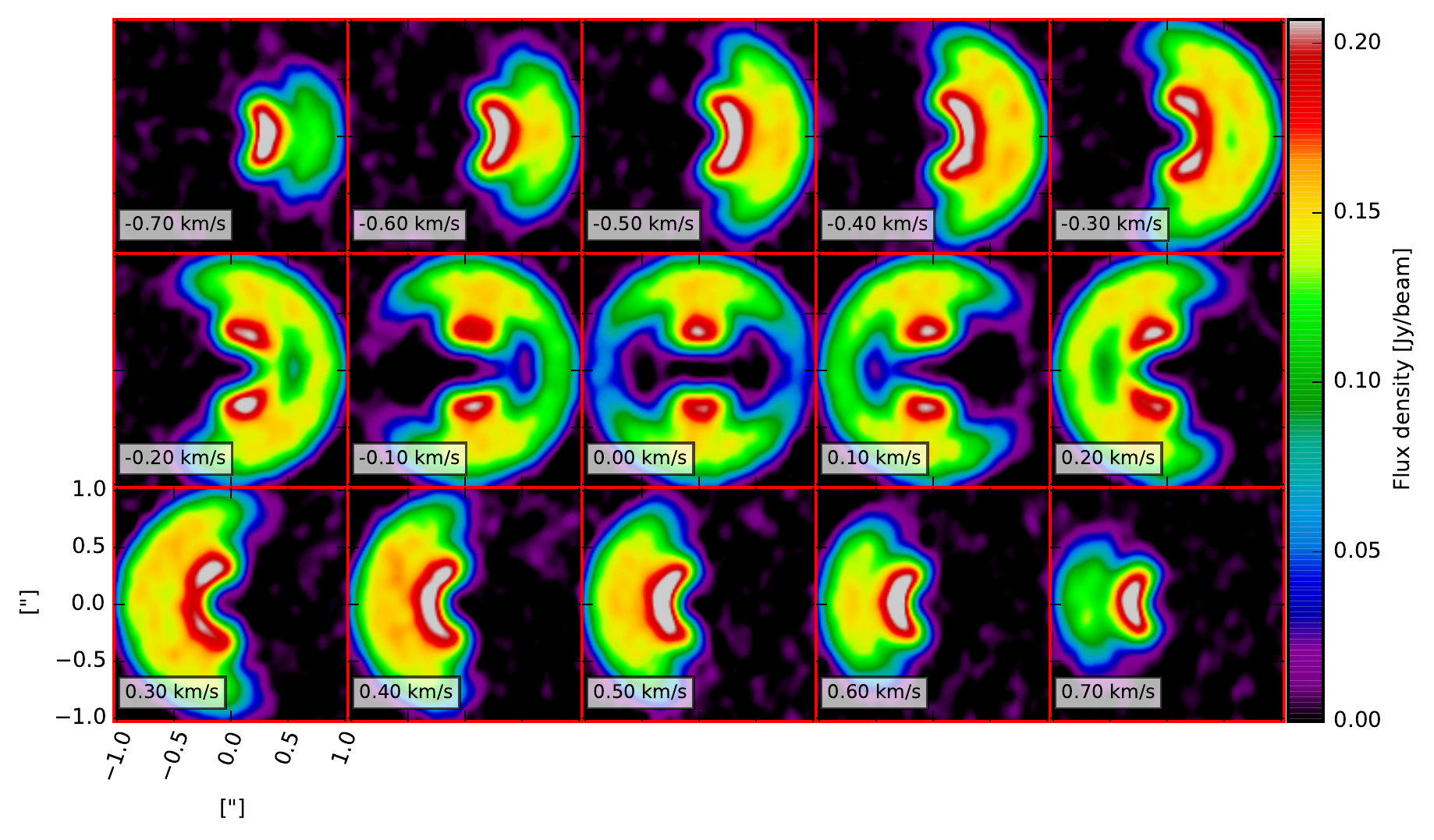}
   \caption{$^{12}$C$^{16}$O~(3-2) velocity-channel map of the same model shown in Fig. \ref{pic:sp1_k05m05st00i0CO3_velo_ch_map_alma14}, but
            with a disk inclination of 10$^{\circ}$. The flux density in Jy/beam is color coded for each individual channel. 
            }\label{pic:sp1_k05m05st00i10CO3_velo_ch_map_alma14}
 \end{figure*}

\begin{figure*}
   \includegraphics[width=\textwidth]{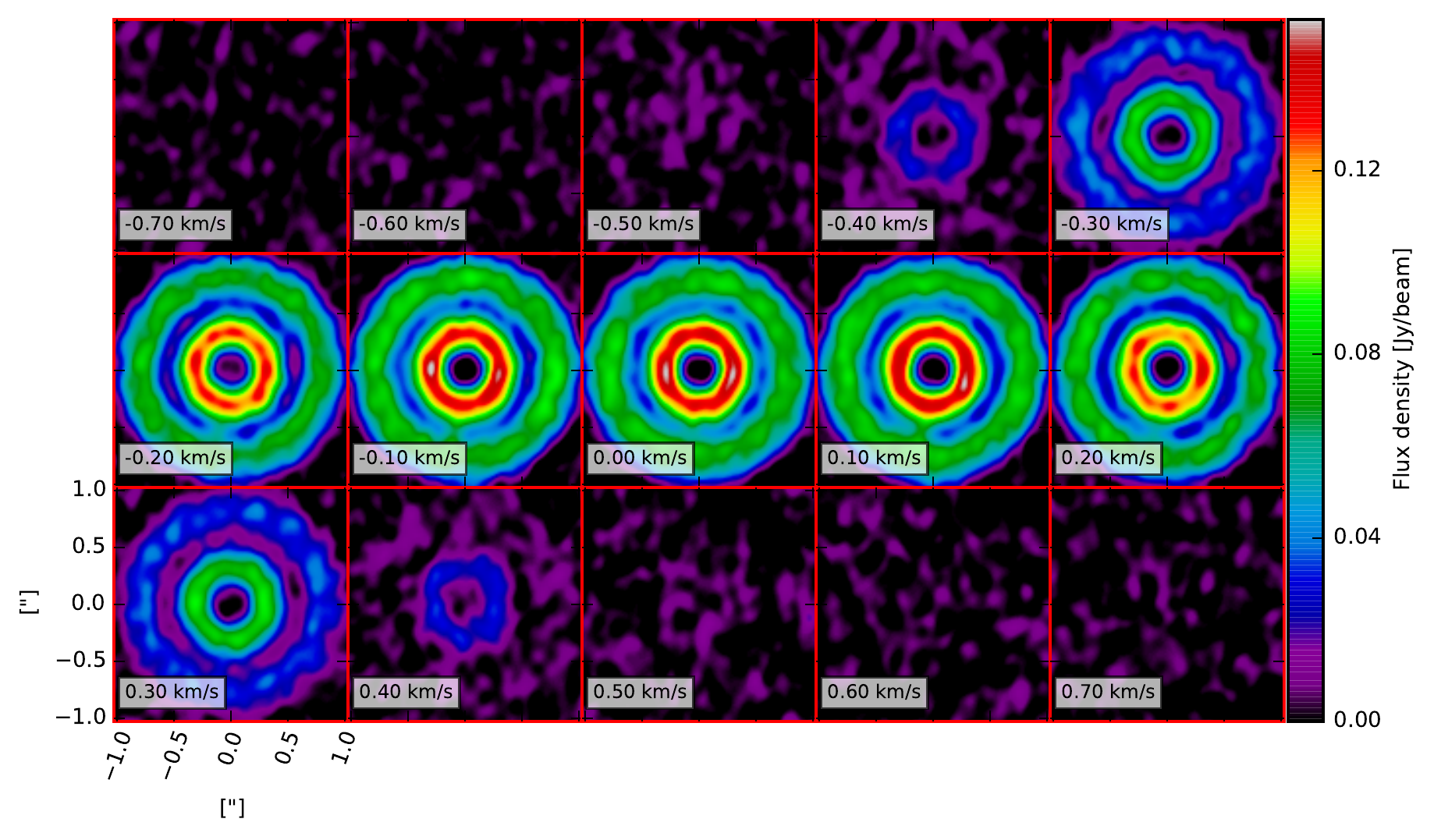}
   \caption{HCO$^+$~(4-3) velocity-channel map of a model with an embedded Jupiter-mass planet as it 
            could be observed using ALMA configuration 14 (1.6 km baseline). 
            The flux density in Jy/beam is color coded for each individual channel. 
            The inclination of the disk amounts to 0$^{\circ}$ (\textit{face-on}). 
            Compared to the same model but simulated $^{12}$C$^{16}$O~(3-2) observation, the line flux is about 35\% lower
            owing to the distinct lower abundance of HCO$^+$.
            However, because of the low abundance the gap is clearly detectable, even at the central wavelength.
            }\label{pic:sp1_k05m05st00i0HCO4_velo_ch_map_alma14}
 \end{figure*}

 In Fig.\ref{pic:sp1_k05m05st00i0CO3_velo_ch_map_alma14} and Fig.\ref{pic:sp1_k05m05st00i10CO3_velo_ch_map_alma14} a face-on
 and an inclined disk model are shown, respectively, as they could be observed with ALMA using configuration 14 ($\sim$ 1.6 km baseline) within three hours
 (corresponding ideal maps are shown in Figs. \ref{pic:sp1_k05m05st00i0CO3_velo_ch_map} and \ref{pic:sp1_k05m05st00i10CO3_velo_ch_map}, respectively).
 More compact baselines result in higher flux densities and consequently higher signal-to-noise ratios, while
 larger baselines allow  gaps of compact disks to be traced at the cost of decreasing signal-to-noise ratio.
 We find for the
 most extended disk configurations (R$_{\text{in}}$ $\sim$ 45~AU, R$_{\text{out}}$ $\sim$~200 AU) in our model setup, maximum baselines of about 1 km
 are sufficient to identify the gap in these disks, but fail for the smaller disks. \\
 For the compact disks in our parameter space, higher spatial resolution and thus 
 longer baselines are required. In our sample the smallest disks have an inner/outer radius of 14~AU/63~AU, respectively. We find that 
 unambiguous gap detection is feasible using intermediate baselines of about 1-3~km. 
 Longer baselines are not required and only result in sparser uv-coverage and thus a lower S/N of the reconstructed maps.
 Furthermore, many of the smaller planetary signatures (inner and outer spiral arms and disk inhomogenities) are eliminated/smoothed out
 because of beam convolution and worse uv-coverage.
 However, for this particular model, and for most of the expanded models in our sample,
 the  gap can be identified in both inclination cases using ALMA configuration 14.
 For the same disk, the resulting HCO$^{+}$~(4-3) velocity-channel map for ALMA configuration 14 is shown in Fig.
 \ref{pic:sp1_k05m05st00i0HCO4_velo_ch_map_alma14}. The HCO$^+$ molecule is less abundant than $^{12}$C$^{16}$O and consequently the line is
 less optically thick. Therefore, regions closer to the disk midplane can be observed and the gap appears more pronounced. 
 In contrast, the line flux itself is about 35\% weaker than the simulated $^{12}$C$^{16}$O~(3-2) line of the same model.
\subsubsection{Gap detection results}
\begin{figure}
   \includegraphics[width=\columnwidth]{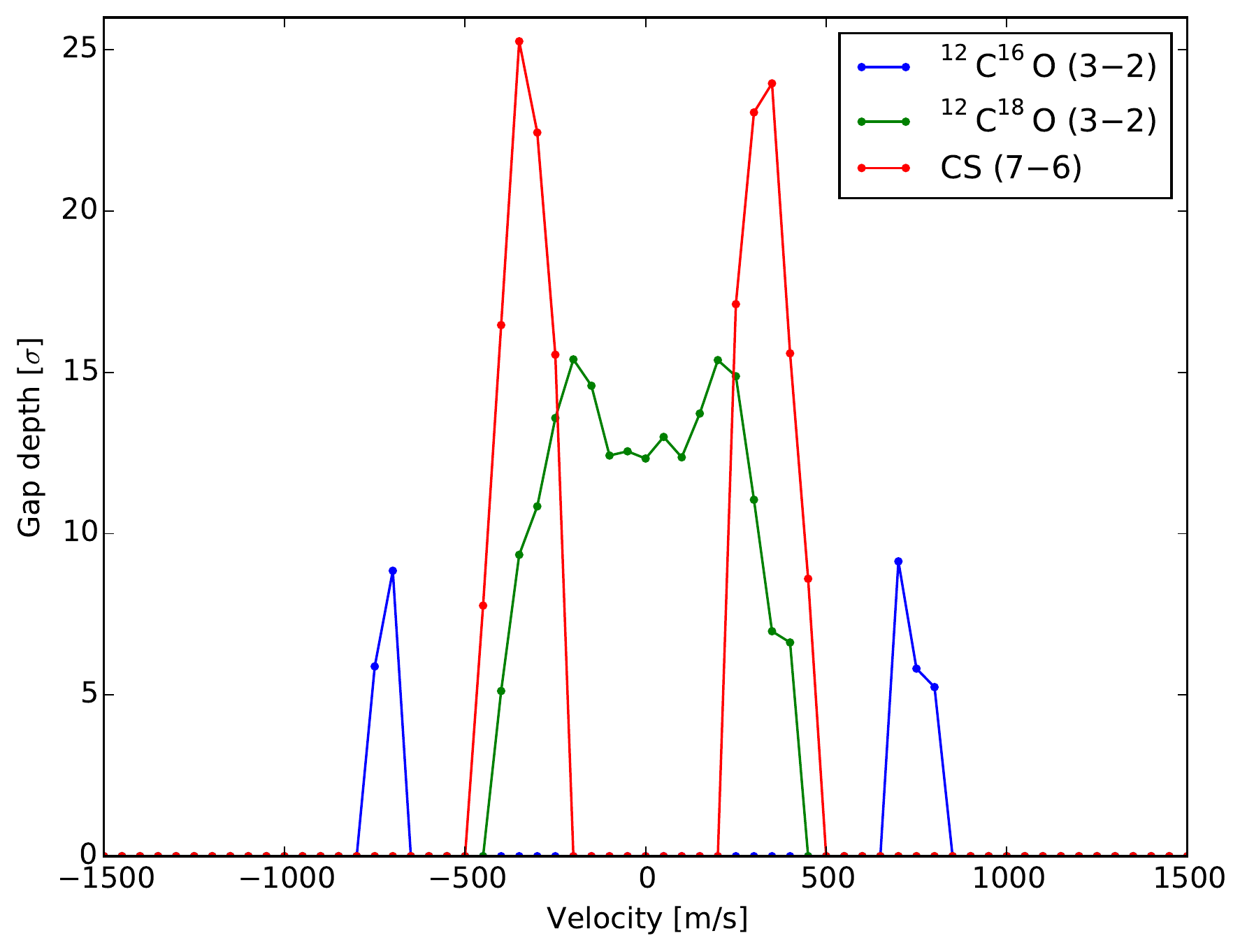}
    \caption{Measured gap depth in units of $\sigma$ over velocity. 
             Gap detection with the optical thick $^{12}$C$^{16}$O~(3-2) transition (blue) is feasible, but only at the line edges (e.g., $\pm$ 700 m/s). 
             In contrast, the less abundant $^{12}$C$^{18}$O~(3-2) transition (green) allows for gap detection at the line center. 
             However, the CS~(7-6) gives the most significant gap depth of about 17$\sigma$ at $\pm$ 350 m/s.
             }\label{pic:gap_detect_thickness}
 \end{figure}

\begin{figure*}       
   \begin{minipage}[b]{0.5\textwidth}
     \includegraphics[width=1\textwidth]{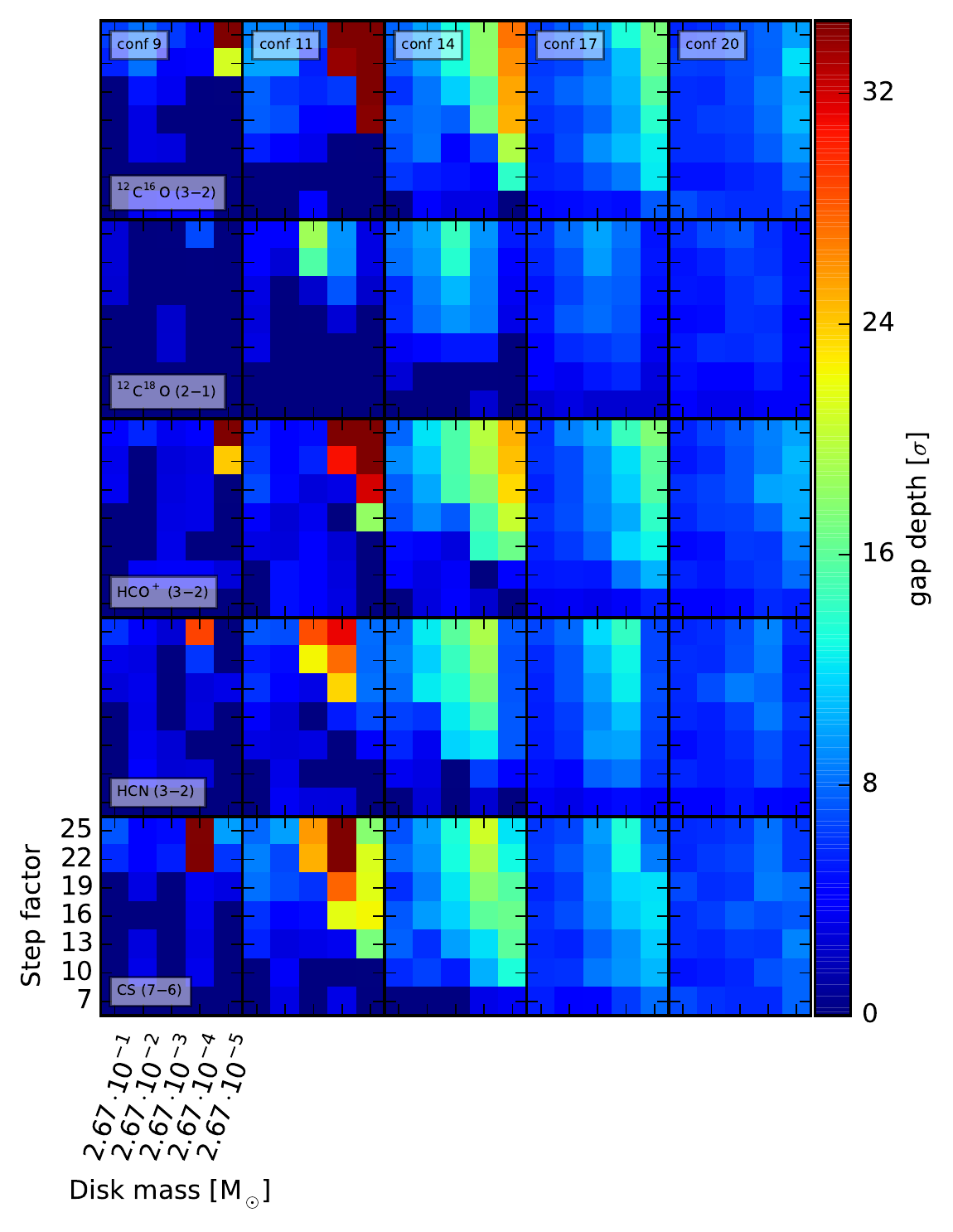}
   \end{minipage}        
   \begin{minipage}[b]{0.5\textwidth}
     \includegraphics[width=1\textwidth]{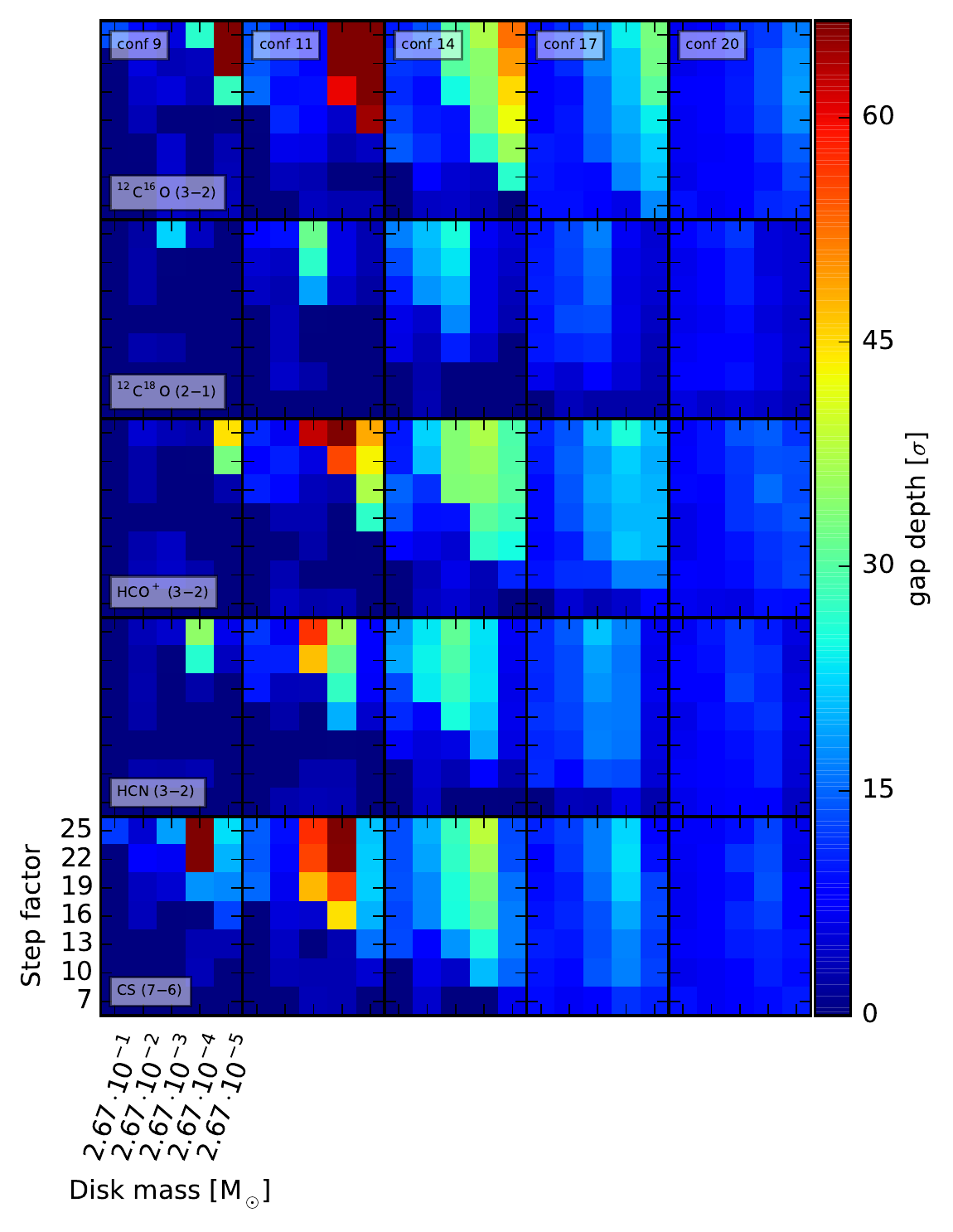}
 \end{minipage}
     \caption{
              Overview maps of the detectability of gaps around Herbig Ae (left) and T~Tauri (right) stars. 
              Each column represents one ALMA configuration and each row represents one
              molecule transition. The x-axis of the inner plot gives the total mass of the disk, and on the y-axis the scaling factor k
              is linearly connected to the spatial dimension of the disks. The depth of the gap is color coded in units of $\sigma$. 
              We only consider a gap as detected when the depth is at least three times the background noise $\sigma$. 
              For both stars, gap detection is feasible for a wide range of models in our setup within three hours of observation. The difference between the two
              stars is mostly due to the much higher luminosity.
              Similar to the results in the ideal case, gap detection is feasible but is restricted to low-mass disks (M$_{\text{disk}} \le 10^{-3}$).
              Gap detection using the most extent configuration (max. baseline $\sim$16~km) is not feasible within three 
              hours of observation, which is mostly due to the small bandwidth of 50~m/s and is not shown in this overview plot.
              }\label{pic:gap_detects}
 \end{figure*}

 Our  main intention is to investigate the feasibility of tracing gaps in molecular lines with ALMA. 
 To allow  this for a large range of disk models, we define a general criterion for a quantitative measure of gap detection in the
 simulated ALMA maps. We apply the same algorithm as in the case of ideal maps (see Sect.~\ref{sec:how_to_gap}).
 The resulting gap depth is given in units of $\sigma$ to clarify the meaning of the gap detection.
 Only if the depth of the gap is greater than three times the noise level $\sigma$ do we consider the gap as detected. \\
 First, we demonstrate that the optical depth effect (see Sect. \ref{sec:individaleffects} for a discussion of the ideal case) has direct observational impact. 
 As shown in the ideal case, the choice of the molecule for a given disk mass/size is most important for 
 unambiguous gap detection. In Fig.~\ref{pic:gap_detect_thickness} we present the gap depth for the
 same transitions as  in Fig.~\ref{pic:k05m03st02i0_gap_best_velo}, but now for a simulated ALMA observation utilizing configuration 14 ($\sim$ 1.6~km
 baseline). Both results are closely comparable as the gap depth in the ideal case shows the same trend as the simulated ALMA observations, i.e., 
 $^{12}$C$^{16}$O~(3-2) cannot be used as a tracer for gaps, and $^{12}$C$^{18}$O~(3-2) allows for gap detection even at the line center, 
 but the CS~(7-6) transition traces the gap best and results in a gap depth of about 17$\sigma$ at $\pm$ 350 m/s.
 \newline
 Second, our results are summarized in the overview charts for the Herbig Ae (left) and T~Tauri (right)
 star (Fig. \ref{pic:gap_detects}).
 Compact ALMA configurations (9, 11) result in high signal-to-noise ratios (S/N), but lack of spatial 
 resolution. As a consequence, utilizing the most compact ALMA configuration nine gaps are only identifiable for the most extended models
 of our sample (e.g., disk models with R$_{\text{out}}$ = 255~AU).
 If we consider longer baselines (higher spatial resolution) in the range of $\sim$1.6~km~-~2.3~km
 we are able to identify gaps for more compact models with a $\sigma$~>~3. 
 Intermediate baselines (e.g., configuration 11-14, $\sim$1.0-1.6~km) potentially allow  gaps to be traced for the 
 wide range of models considered in this study.\\
 The comparison between the molecules considered shows that CS, HCN, and HCO$^{+}$ allow
 the detection of gaps even for disks with intermediate masses ($\sim$10$^{-3}$~M$_{\odot}$), while $^{12}$C$^{16}$O tends to fail.  
 As already discussed in Sect. \ref{sec:para_lineRT}, owing to the optical thick $^{12}$C$^{16}$O line only the surface of the disks can be observed, 
 which is barely perturbed by the planet-disk interaction.
 In contrast, for lines of molecules with lower abundances
 the surface is still optically thin even for more massive disks. We find that especially 
 CS and HCN are proper candidates to serve as a tracer for gap detection in the mass range from $\sim$10$^{-3}$~M$_{\odot}$ to 10$^{-4}$~M$_{\odot}$.\\ 
 For more massive disks
 ($M_{\text{disk}}\geq 10^{-2}$~M$_{\odot}$) the $^{12}$C$^{16}$O isotopologue $^{12}$C$^{18}$O is an adequate candidate. 
 However, owing to the resulting low line flux gaps these massive disks should be observed in the continuum 
 because in the submm wavelength range provided by ALMA the disk is still optically thin in the gap region \citep[see][]{Ruge2013}.\\
 Finally, we find that because of the assumed observation time of 3~h, long-baseline
 configurations (20, 28) are not suited to identifying gaps because of the low intrinsic flux and inadequate uv-coverage.
  \section{Summary}\label{sec:conclusion}
Our intention was to explore how planet-induced structures, in particular planetary gaps, 
appear in high-angular resolution molecular line maps, and subsequently if ALMA
would allow the detection of these structures.
For this reason, we developed the line radiative transfer code called \textit{Mol3D}. It allows the calculation of 
the level-populations with several N-LTE approximate methods and the production of ideal velocity-channel maps.
First, we tested the new code against other established (molecular line) RT codes.\\
Based on 3D hydrodynamic simulations carried out with the PLUTO code, we simulated synthetic velocity-channel maps for 
five molecules ($^{12}$C$^{16}$O, $^{12}$C$^{18}$O, HCO$^{+}$, HCN, and CS) for a total of 32 rotational transitions. 
We summarize our findings:
\begin{itemize}
 \item The patterns of the global (Keplerian) velocity field in the corresponding channel maps of inclined disks mask planet-induced structures.
       Gap identification can be ambiguous in these cases.
 \item Molecules that tend to become optically thick (like $^{12}$C$^{16}$O) only trace the upper layers of the disk, which are rarely affected
       by the planet-disk interaction. In consequence, gap detection for massive disks ($\gtrsim 10^{-2}$ M$_{\odot}$) is barely feasible with molecular lines.
 \item For lower mass disks, depending on the molecule and its optical properties (abundances, excitation state) 
       the gap can be observed for a broad range of considered disk configurations,  not necessarily at
       the central wavelength of this transition but at the line wings. Because of the higher optical depth of the line at its center, the $\tau$ = 1 surface
       traces only the upper unperturbed disk layers, while the $\tau$ = 1 surface of the line wings traces deeper and thus more strongly perturbed layers.    
 \item With ALMA it is feasible to identify gaps using molecular lines for several disk configurations, if one respects the effects we outlined above 
       (i.e., by choosing an appropriate optically thin molecule and corresponding transition). 
       In particular, for low-mass disks ($\lesssim 10^{-3}$~M$_{\odot}$) the feasibility of identifying gaps using molecular lines is very promising. For 
       this mass range molecular line observations fill the discovery space for gaps where continuum 
       observations are supposed to fail \citep[see Fig. 8 in the complementary study of][]{Ruge2013}.        
 \item Our simulations are based on an assumed distance of 140~pc and a fixed sky coordinate 
       (max. elevation $\sim45^{\circ}$) to give reliable results for the Taurus star-forming region. Thus, for many disks 
       located closer or at more favorable positions on the sky the feasibility of detecting gaps is expected to be even higher.  
\end{itemize}
We emphasize, that the observation of molecules
with different abundances allow for unambiguous identification of gaps and the physical and chemical composition of protoplanetary disks. Especially
for low-mass disks where gap detection in the continuum is difficult owing to the low intrinsic flux, molecular line observations are a promising alternative
with which to identify gaps.\\
Finally, we note that our study is based on HD simulations. In addition, \citet{Uribe2011} have remarked that magnetohydrodynamics (MHD) simulations 
result in more extended gaps owing to the presence of the weak magnetic field. Including intrinsic stellar radiation, the form of the gaps also changes  
because the gap walls can be directly radiated by the host star, resulting in deeper gaps. 
This effect is even higher for the outer wall \citep{Jang-Condell2012}.
Using smoothed particle hydrodynamic (SPH) simulations, \citet{Gonzalez2012} found that gaps are deeper for larger dust grains and bigger planets. 
In addition, we only considered a Jupiter-mass planet in
this study. More massive planets would enlarge the gap width owing to the higher gravitational potential.
If these effects are present it would therefore be even easier to observe gaps with ALMA than proposed within this study.
  \begin{acknowledgements}
 This presented work is funded through the DFG grant: WO 857/10-1.
 Special thanks go to Yaroslav Pavlyuchenkov for his comments and advice, which were very helpful in the early development phase of
 \textit{Mol3D}. We also want to acknowledge Peter Scicluna for language editing and Jan Philipp Ruge for the helpful advice, 
 especially regarding the use of the CASA simulator.
  \end{acknowledgements}
  \bibliographystyle{aa}
        \bibliography{bib_astro-ph}

\begin{thebibliography}{75}
\expandafter\ifx\csname natexlab\endcsname\relax\def\natexlab#1{#1}\fi

\bibitem[{Aikawa \& Nomura(2006)}]{Aikawa2006}
Aikawa, Y. \& Nomura, H. 2006, The Astrophysical Journal, 642, 1152

\bibitem[{Andrews {et~al.}(2011)Andrews, Wilner, Espaillat, Hughes, Dullemond,
  {McClure}, Qi, \& Brown}]{Andrews2011}
Andrews, S.~M., Wilner, D.~J., Espaillat, C., {et~al.} 2011, The Astrophysical
  Journal, 732, 42

\bibitem[{Andrews {et~al.}(2010)Andrews, Wilner, Hughes, Qi, \&
  Dullemond}]{Andrews2010}
Andrews, S.~M., Wilner, D.~J., Hughes, A.~M., Qi, C., \& Dullemond, C.~P. 2010,
  The Astrophysical Journal, 723, 1241

\bibitem[{Bjorkman \& Wood(2001)}]{Bjorkman2001}
Bjorkman, J.~E. \& Wood, K. 2001, The Astrophysical Journal, 554, 615

\bibitem[{Bruderer {et~al.}(2014)Bruderer, van~der Marel, van Dishoeck, \& van
  Kempen}]{Bruderer2014}
Bruderer, S., van~der Marel, N., van Dishoeck, E.~F., \& van Kempen, T.~A.
  2014, Astronomy and Astrophysics, 562, 26

\bibitem[{Casassus {et~al.}(2013)Casassus, van~der Plas, M, Dent, Fomalont,
  Hagelberg, Hales, Jord{\'a}n, Mawet, M{\'e}nard, Wootten, Wilner, Hughes,
  Schreiber, Girard, Ercolano, Canovas, Rom{\'a}n, \& Salinas}]{Casassus2013a}
Casassus, S., van~der Plas, G., M, S.~P., {et~al.} 2013, Nature, 493, 191

\bibitem[{Castro-Carrizo {et~al.}(2007)Castro-Carrizo, Quintana-Lacaci,
  Bujarrabal, Neri, \& Alcolea}]{Castro-Carrizo2007}
Castro-Carrizo, A., Quintana-Lacaci, G., Bujarrabal, V., Neri, R., \& Alcolea,
  J. 2007, Astronomy and Astrophysics, 465, 457

\bibitem[{Chapillon {et~al.}(2012{\natexlab{a}})Chapillon, Dutrey, Guilloteau,
  Pi{\'e}tu, Wakelam, Hersant, Gueth, Henning, Launhardt, Schreyer, \&
  Semenov}]{Chapillon2012a}
Chapillon, E., Dutrey, A., Guilloteau, S., {et~al.} 2012{\natexlab{a}}, The
  Astrophysical Journal, 756, 58

\bibitem[{Chapillon {et~al.}(2012{\natexlab{b}})Chapillon, Guilloteau, Dutrey,
  Pi{\'e}tu, \& Gu{\'e}lin}]{Chapillon2012}
Chapillon, E., Guilloteau, S., Dutrey, A., Pi{\'e}tu, V., \& Gu{\'e}lin, M.
  2012{\natexlab{b}}, Astronomy and Astrophysics, 537, A60

\bibitem[{Cleeves {et~al.}(2011)Cleeves, Bergin, Bethell, Calvet, Fogel,
  Sauter, \& Wolf}]{Cleeves2011}
Cleeves, L.~I., Bergin, E.~A., Bethell, T.~J., {et~al.} 2011, The Astrophysical
  Journal Letters, 743, L2

\bibitem[{D'Alessio {et~al.}(2005)D'Alessio, Hartmann, Calvet,
  Franco-Hern{\'a}ndez, Forrest, Sargent, Furlan, Uchida, Green, Watson, Chen,
  Kemper, Sloan, \& Najita}]{DAlessio2005}
D'Alessio, P., Hartmann, L., Calvet, N., {et~al.} 2005, {ApJ}, 621, 461

\bibitem[{Dartois {et~al.}(2003)Dartois, Dutrey, \& Guilloteau}]{Dartois2003}
Dartois, E., Dutrey, A., \& Guilloteau, S. 2003, Astronomy and Astrophysics,
  399, 773

\bibitem[{de~Gregorio-Monsalvo {et~al.}(2013)de~Gregorio-Monsalvo, M{\'e}nard,
  Dent, Pinte, L{\'o}pez, Klaassen, Hales, Cort{\'e}s, Rawlings, Tachihara,
  Testi, Takahashi, Chapillon, Mathews, Juhasz, Akiyama, Higuchi, Saito, Nyman,
  Phillips, Rod{\'o}n, Corder, \& Van~Kempen}]{deGregorio-Monsalvo2013}
de~Gregorio-Monsalvo, I., M{\'e}nard, F., Dent, W., {et~al.} 2013, Astronomy
  and Astrophysics, 557, 133

\bibitem[{de~Jong {et~al.}(1980)de~Jong, Boland, \& Dalgarno}]{deJong1980}
de~Jong, T., Boland, W., \& Dalgarno, A. 1980, Astronomy and Astrophysics, 91,
  68

\bibitem[{de~Jong {et~al.}(1975)de~Jong, Dalgarno, \& Chu}]{deJong1975}
de~Jong, T., Dalgarno, A., \& Chu, S.-I. 1975, The Astrophysical Journal, 199,
  69

\bibitem[{Dohnanyi(1969)}]{Dohnanyi1969}
Dohnanyi, J.~S. 1969, Journal of Geophysical Research, 74, 2531

\bibitem[{Dutrey {et~al.}(2008)Dutrey, Guilloteau, Pi{\'e}tu, Chapillon, Gueth,
  Henning, Launhardt, Pavlyuchenkov, Schreyer, \& Semenov}]{Dutrey2008}
Dutrey, A., Guilloteau, S., Pi{\'e}tu, V., {et~al.} 2008, Astronomy and
  Astrophysics, 490, L15

\bibitem[{Flock {et~al.}(2015)Flock, Ruge, Dzyurkevich, Henning, Klahr, \&
  Wolf}]{Flock2015}
Flock, M., Ruge, J.~P., Dzyurkevich, N., {et~al.} 2015, Astronomy and
  Astrophysics, 574, A68

\bibitem[{Forrest {et~al.}(2004)Forrest, Sargent, Furlan, D'Alessio, Calvet,
  Hartmann, Uchida, Green, Watson, Chen, Kemper, Keller, Sloan, Herter, Brandl,
  Houck, Barry, Hall, Morris, Najita, \& Myers}]{Forrest2004}
Forrest, W.~J., Sargent, B., Furlan, E., {et~al.} 2004, {ApJS}, 154, 443

\bibitem[{Goldreich \& Tremaine(1980)}]{Goldreich1980}
Goldreich, P. \& Tremaine, S. 1980, The Astrophysical Journal, 241, 425

\bibitem[{Gonzalez {et~al.}(2012)Gonzalez, Pinte, Maddison, M{\'e}nard, \&
  Fouchet}]{Gonzalez2012}
Gonzalez, J.-F., Pinte, C., Maddison, S.~T., M{\'e}nard, F., \& Fouchet, L.
  2012, Astronomy and Astrophysics, 547, A58

\bibitem[{Goto {et~al.}(2012)Goto, van~der Plas, van~den Ancker, Dullemond,
  Carmona, Henning, Meeus, Linz, \& Stecklum}]{Goto2012}
Goto, M., van~der Plas, G., van~den Ancker, M., {et~al.} 2012, Astronomy and
  Astrophysics, 539, 81

\bibitem[{Gr{\"a}fe \& Wolf(2013)}]{Grafe2013a}
Gr{\"a}fe, C. \& Wolf, S. 2013, Astronomy \& Astrophysics, 552, A88

\bibitem[{Gr{\"a}fe {et~al.}(2013)Gr{\"a}fe, Wolf, Guilloteau, Dutrey,
  Stapelfeldt, Pontoppidan, \& Sauter}]{Grafe2013}
Gr{\"a}fe, C., Wolf, S., Guilloteau, S., {et~al.} 2013, Astronomy \&
  Astrophysics, 553, A69

\bibitem[{Hersant {et~al.}(2009)Hersant, Wakelam, Dutrey, Guilloteau, \&
  Herbst}]{Hersant2009}
Hersant, F., Wakelam, V., Dutrey, A., Guilloteau, S., \& Herbst, E. 2009,
  Astronomy and Astrophysics, 493, L49

\bibitem[{H{\"o}gbom(1974)}]{Hogbom1974}
H{\"o}gbom, J.~A. 1974, Astronomy and Astrophysics Supplement Series, 15, 417

\bibitem[{Hogerheijde \& van~der Tak(2000)}]{Hogerheijde2000}
Hogerheijde, M.~R. \& van~der Tak, F. F.~S. 2000, Astronomy and Astrophysics,
  362, 697

\bibitem[{Hughes {et~al.}(2011)Hughes, Wilner, Andrews, Qi, \&
  Hogerheijde}]{Hughes2011}
Hughes, A.~M., Wilner, D.~J., Andrews, S.~M., Qi, C., \& Hogerheijde, M.~R.
  2011, The Astrophysical Journal, 727, 85

\bibitem[{Jang-Condell \& Turner(2012)}]{Jang-Condell2012}
Jang-Condell, H. \& Turner, N.~J. 2012, The Astrophysical Journal, 749, 153

\bibitem[{Kastner {et~al.}(1997)Kastner, Zuckerman, Weintraub, \&
  Forveille}]{Kastner1997}
Kastner, J.~H., Zuckerman, B., Weintraub, D.~A., \& Forveille, T. 1997,
  Science, 277, 67

\bibitem[{Kennedy {et~al.}(2014)Kennedy, Murphy, Lisse, M{\'e}nard, Sitko,
  Wyatt, Bayliss, {DeMeo}, Crawford, Kim, Rudy, Russell, Sibthorpe, Skinner, \&
  Zhou}]{Kennedy2014}
Kennedy, G.~M., Murphy, S.~J., Lisse, C.~M., {et~al.} 2014, Monthly Notices of
  the Royal Astronomical Society, 438, 3299

\bibitem[{Lucy(1999)}]{Lucy1999}
Lucy, L.~B. 1999, Astronomy and Astrophysics, 344, 282

\bibitem[{Madlener {et~al.}(2012)Madlener, Wolf, Dutrey, \&
  Guilloteau}]{Madlener2012}
Madlener, D., Wolf, S., Dutrey, A., \& Guilloteau, S. 2012, Astronomy \&
  Astrophysics, 543, A81

\bibitem[{Mathis {et~al.}(1977)Mathis, Rumpl, \& Nordsieck}]{Mathis1977}
Mathis, J.~S., Rumpl, W., \& Nordsieck, K.~H. 1977, The Astrophysical Journal,
  217, 425

\bibitem[{Mignone {et~al.}(2007)Mignone, Bodo, Massaglia, Matsakos, Tesileanu,
  Zanni, \& Ferrari}]{Mignone2007}
Mignone, A., Bodo, G., Massaglia, S., {et~al.} 2007, The Astrophysical Journal
  Supplement Series, 170, 228

\bibitem[{Mihalas {et~al.}(1978)Mihalas, Auer, \& Mihalas}]{Mihalas1978}
Mihalas, D., Auer, L.~H., \& Mihalas, B.~R. 1978, The Astrophysical Journal,
  220, 1001

\bibitem[{Millar {et~al.}(1997)Millar, Farquhar, \& Willacy}]{Millar1997}
Millar, T.~J., Farquhar, P. R.~A., \& Willacy, K. 1997, Astronomy and
  Astrophysics Supplement Series, 121, 139

\bibitem[{Mordasini {et~al.}(2010)Mordasini, Klahr, Alibert, Benz, \&
  Dittkrist}]{Mordasini2010}
Mordasini, C., Klahr, H., Alibert, Y., Benz, W., \& Dittkrist, K.-M. 2010,
  proceedings workshop "Circumstellar disks and planets: Science cases for the
  second generation {VLTI} instrumentation"

\bibitem[{{\"O}berg {et~al.}(2010){\"O}berg, Qi, Fogel, Bergin, Andrews,
  Espaillat, van Kempen, Wilner, \& Pascucci}]{Oberg2010}
{\"O}berg, K.~I., Qi, C., Fogel, J. K.~J., {et~al.} 2010, The Astrophysical
  Journal, 720, 480

\bibitem[{Osterbrock \& Ferland(2006)}]{Osterbrock2006}
Osterbrock, D.~E. \& Ferland, G.~J. 2006, Astrophysics of gaseous nebulae and
  active galactic nuclei (Palgrave Macmillan)

\bibitem[{Paardekooper \& Mellema(2004)}]{Paardekooper2004}
Paardekooper, S.-J. \& Mellema, G. 2004, Astronomy and Astrophysics, 425, L9

\bibitem[{Paardekooper \& Mellema(2006)}]{Paardekooper2006}
Paardekooper, S.-J. \& Mellema, G. 2006, Astronomy \& Astrophysics, 453, 12

\bibitem[{Papaloizou \& Lin(1984)}]{Papaloizou1984}
Papaloizou, J. \& Lin, D. N.~C. 1984, The Astrophysical Journal, 285, 818

\bibitem[{Pascucci {et~al.}(2004)Pascucci, Wolf, Steinacker, Dullemond,
  Henning, Niccolini, Woitke, \& Lopez}]{Pascucci2004}
Pascucci, I., Wolf, S., Steinacker, J., {et~al.} 2004, Astronomy and
  Astrophysics, 417, 793

\bibitem[{Pavlyuchenkov {et~al.}(2007)Pavlyuchenkov, Semenov, Henning,
  Pi{\'e}tu, Launhardt, \& Dutrey}]{Pavlyuchenkov2007}
Pavlyuchenkov, Y., Semenov, D., Henning, T., {et~al.} 2007, The Astrophysical
  Journal, 669, 1262

\bibitem[{Pavlyuchenkov {et~al.}(2008)Pavlyuchenkov, Wiebe, Shustov, Henning,
  Launhardt, \& Semenov}]{Pavlyuchenkov2008}
Pavlyuchenkov, Y., Wiebe, D., Shustov, B., {et~al.} 2008, The Astrophysical
  Journal, 689, 335

\bibitem[{Pavlyuchenkov \& Shustov(2004)}]{Pavlyuchenkov2004}
Pavlyuchenkov, Y.~N. \& Shustov, B.~M. 2004, Astronomy Reports, 48, 315

\bibitem[{Petry \& {CASA Development Team}(2012)}]{Petry2012}
Petry, D. \& {CASA Development Team}. 2012, Astronomical Data Analysis Software
  and Systems {XXI}, 461, 849

\bibitem[{Pi{\'e}tu {et~al.}(2007)Pi{\'e}tu, Dutrey, \& Guilloteau}]{Pietu2007}
Pi{\'e}tu, V., Dutrey, A., \& Guilloteau, S. 2007, Astronomy and Astrophysics,
  467, 163

\bibitem[{Qi {et~al.}(2013)Qi, {\"O}berg, Wilner, \& Rosenfeld}]{Qi2013}
Qi, C., {\"O}berg, K.~I., Wilner, D.~J., \& Rosenfeld, K.~A. 2013, The
  Astrophysical Journal Letters, 765, L14

\bibitem[{Rosenfeld {et~al.}(2013)Rosenfeld, Andrews, Hughes, Wilner, \&
  Qi}]{Rosenfeld2013}
Rosenfeld, K.~A., Andrews, S.~M., Hughes, A.~M., Wilner, D.~J., \& Qi, C. 2013,
  The Astrophysical Journal, 774, 16

\bibitem[{Ruge {et~al.}(2013)Ruge, Wolf, Uribe, \& Klahr}]{Ruge2013}
Ruge, J.~P., Wolf, S., Uribe, A.~L., \& Klahr, H.~H. 2013, Astronomy \&
  Astrophysics, 549, A97

\bibitem[{Ruge {et~al.}(2014)Ruge, Wolf, Uribe, \& Klahr}]{Ruge2014}
Ruge, J.~P., Wolf, S., Uribe, A.~L., \& Klahr, H.~H. 2014, Astronomy and
  Astrophysics, 572, L2

\bibitem[{Rybicki \& Hummer(1991)}]{Rybicki1991}
Rybicki, G.~B. \& Hummer, D.~G. 1991, Astronomy and Astrophysics, 245, 171

\bibitem[{Sauter \& Wolf(2011)}]{Sauter2011}
Sauter, J. \& Wolf, S. 2011, Astronomy \& Astrophysics, 527, A27

\bibitem[{Schegerer {et~al.}(2008)Schegerer, Wolf, Ratzka, \&
  Leinert}]{Schegerer2008}
Schegerer, A.~A., Wolf, S., Ratzka, T., \& Leinert, C. 2008, Astronomy and
  Astrophysics, 478, 779

\bibitem[{Sch{\"o}ier {et~al.}(2005)Sch{\"o}ier, van~der Tak, van Dishoeck, \&
  Black}]{Schoier2005}
Sch{\"o}ier, F.~L., van~der Tak, F. F.~S., van Dishoeck, E.~F., \& Black, J.~H.
  2005, Astronomy and Astrophysics, 432, 369

\bibitem[{Semenov {et~al.}(2004)Semenov, Pavlyuchenkov, Henning, Herbst, \& van
  Dishoeck}]{Semenov2004}
Semenov, D., Pavlyuchenkov, Y., Henning, T., Herbst, E., \& van Dishoeck, E.
  2004, Baltic Astronomy, 13, 454

\bibitem[{Semenov {et~al.}(2008)Semenov, Pavlyuchenkov, Henning, Wolf, \&
  Launhardt}]{Semenov2008}
Semenov, D., Pavlyuchenkov, Y., Henning, T., Wolf, S., \& Launhardt, R. 2008,
  The Astrophysical Journal Letters, 673, L195

\bibitem[{Semenov {et~al.}(2005)Semenov, Pavlyuchenkov, Schreyer, Henning,
  Dullemond, \& Bacmann}]{Semenov2005}
Semenov, D., Pavlyuchenkov, Y., Schreyer, K., {et~al.} 2005, The Astrophysical
  Journal, 621, 853

\bibitem[{Semenov {et~al.}(2006)Semenov, Wiebe, \& Henning}]{Semenov2006}
Semenov, D., Wiebe, D., \& Henning, T. 2006, The Astrophysical Journal Letters,
  647, L57

\bibitem[{Shakura \& Sunyaev(1973)}]{Shakura1973}
Shakura, N.~I. \& Sunyaev, R.~A. 1973, Astronomy and Astrophysics, 24, 337

\bibitem[{Tanaka {et~al.}(2002)Tanaka, Takeuchi, \& Ward}]{Tanaka2002}
Tanaka, H., Takeuchi, T., \& Ward, W.~R. 2002, The Astrophysical Journal, 565,
  1257

\bibitem[{Thi {et~al.}(2004)Thi, van Zadelhoff, \& van Dishoeck}]{Thi2004}
Thi, W.-F., van Zadelhoff, G.-J., \& van Dishoeck, E.~F. 2004, Astronomy and
  Astrophysics, 425, 955

\bibitem[{Toro(2009)}]{Toro2009}
Toro, E.~F. 2009, Riemann Solvers and Numerical Methods for Fluid Dynamics
  (Berlin, Heidelberg: Springer Berlin Heidelberg)

\bibitem[{Uribe {et~al.}(2011)Uribe, Klahr, Flock, \& Henning}]{Uribe2011}
Uribe, A.~L., Klahr, H., Flock, M., \& Henning, T. 2011, The Astrophysical
  Journal, 736, 85

\bibitem[{van~der Tak {et~al.}(2007)van~der Tak, Black, Sch{\"o}ier, Jansen, \&
  van Dishoeck}]{vanderTak2007}
van~der Tak, F. F.~S., Black, J.~H., Sch{\"o}ier, F.~L., Jansen, D.~J., \& van
  Dishoeck, E.~F. 2007, Astronomy and Astrophysics, 468, 627

\bibitem[{van Zadelhoff {et~al.}(2002)van Zadelhoff, Dullemond, van~der Tak,
  Yates, Doty, Ossenkopf, Hogerheijde, Juvela, Wiesemeyer, \&
  Sch{\"o}ier}]{vanZadelhoff2002}
van Zadelhoff, G.-J., Dullemond, C.~P., van~der Tak, F. F.~S., {et~al.} 2002,
  Astronomy and Astrophysics, 395, 373

\bibitem[{Ward(1997)}]{Ward1997}
Ward, W.~R. 1997, The Astrophysical Journal Letters, 482, L211

\bibitem[{Weingartner \& Draine(2001)}]{Weingartner2001}
Weingartner, J.~C. \& Draine, B.~T. 2001, The Astrophysical Journal, 548, 296

\bibitem[{Wei{\ss} {et~al.}(2005)Wei{\ss}, Walter, \& Scoville}]{Weis2005}
Wei{\ss}, A., Walter, F., \& Scoville, N.~Z. 2005, Astronomy and Astrophysics,
  438, 533

\bibitem[{Wolf(2003)}]{Wolf2003c}
Wolf, S. 2003, Computer Physics Communications, 150, 99

\bibitem[{Wolf \& D'Angelo(2005)}]{Wolf2005}
Wolf, S. \& D'Angelo, G. 2005, The Astrophysical Journal, 619, 1114

\bibitem[{Wolf {et~al.}(1999)Wolf, Henning, \& Stecklum}]{Wolf1999}
Wolf, S., Henning, T., \& Stecklum, B. 1999, Astronomy and Astrophysics, 349,
  839

\bibitem[{Wolf {et~al.}(2003)Wolf, Padgett, \& Stapelfeldt}]{Wolf2003d}
Wolf, S., Padgett, D.~L., \& Stapelfeldt, K.~R. 2003, The Astrophysical
  Journal, 588, 373

\end{thebibliography}
  \appendix
  \section{Gap depth - complete overview charts}\label{sec:appendix_study}
\begin{figure*}
   \includegraphics[width=\textwidth]{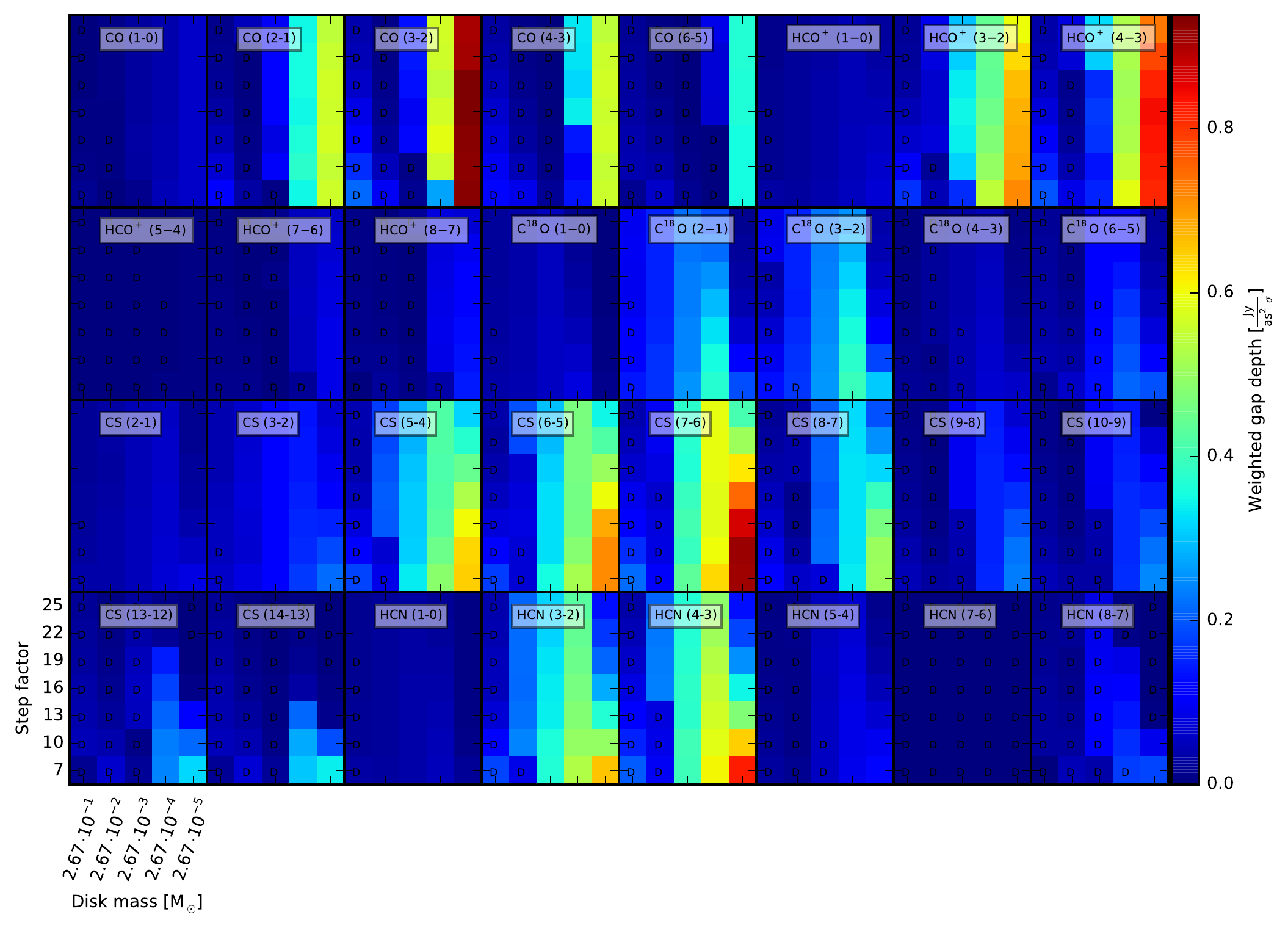}
    \caption{Overview plot of gap depth in the ideal velocity-channel maps for five different molecules. 
             The central radiation source is a TTauri star. The $\sigma$-weighted gap depth is color coded.
             }\label{pic:gap_detect_ideal_s00}
 \end{figure*}

 The full overview charts for all molecules considered in our study are summarized in 
 Fig. \ref{pic:gap_detect_ideal_s00} for the T Tauri case and in Fig. \ref{pic:gap_detect_ideal_s00} for the Herbig Ae case. 
 The derived $\sigma$-weighted gap depth is color coded for each considered disk configuration. 
 \begin{figure*}
   \includegraphics[width=\textwidth]{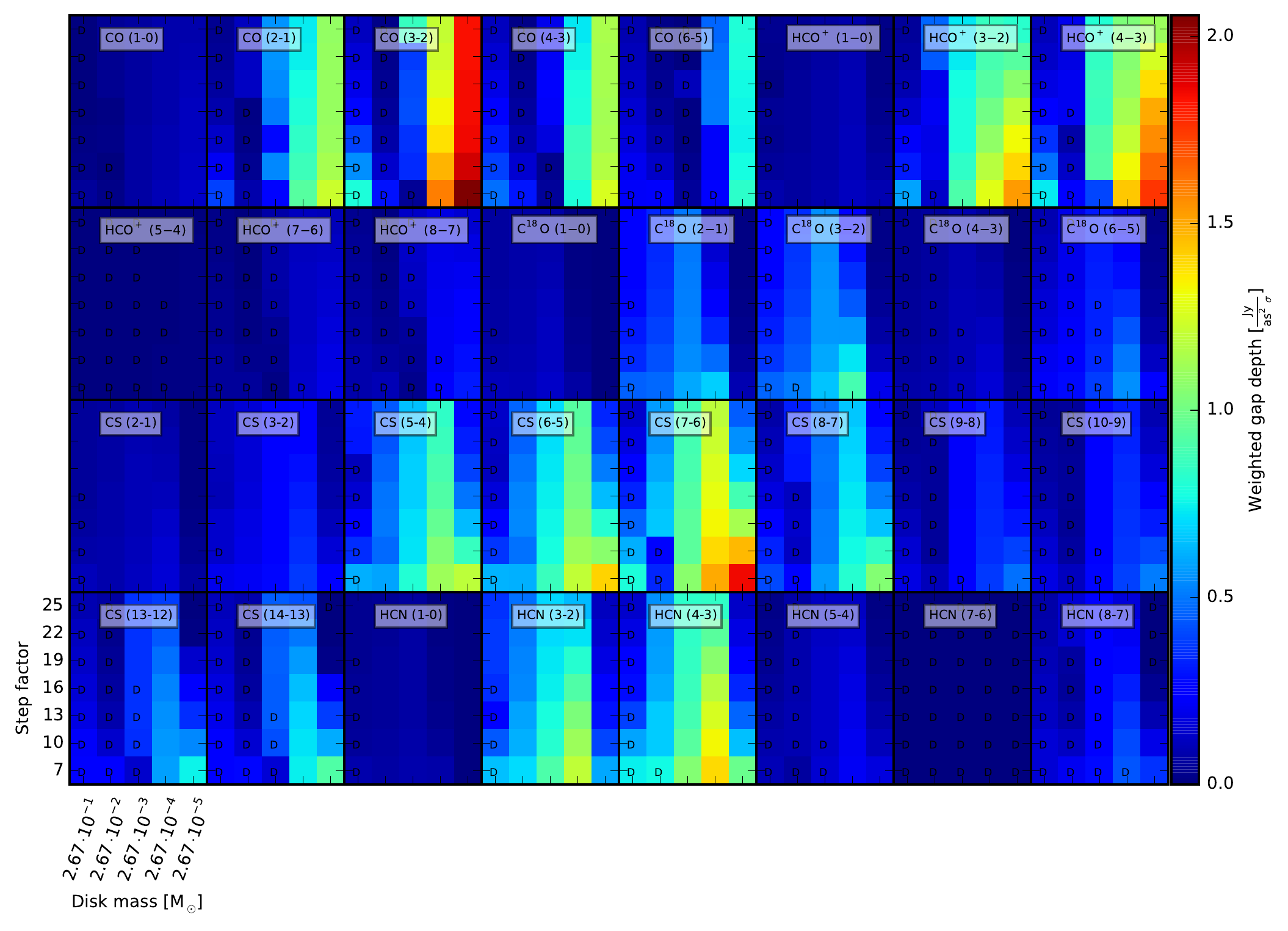}
    \caption{Overview plot of gap depth in the ideal velocity-channel maps for five different molecules. 
             The central radiation source is a Herbig Ae star. The $\sigma$-weighted gap depth is color coded.
             }\label{pic:gap_detect_ideal_s02}
 \end{figure*}

  \section{Mol3D}\label{sec:appendix}
In this section we present a detailed derivation of the underlying methods, physics, and assumptions used by our new 
N-LTE line and dust continuum radiative transfer code \textit{Mol3D}.
\subsection{Radiative transfer equations}
 We start with the transfer equation in the following form,
 \begin{align}
   \frac{\text{d}I_{\nu}(s)} {\text{d}s} = -\alpha_\nu(s) I_{\nu}(s) + j_\nu(s), \label{eq:RTE}
 \end{align}
  with the monochromatic intensity $I_{\nu}$, the emission factor $j_\nu(s)$,  and the total absorption factor $\alpha_\nu(s)$. 
  The intensity from point p$_1$ to p$_2$ along its path
  \textbf{s} can be calculated using the integral form of equation \ref{eq:RTE},
  \begin{align}
   I_{\nu}(\text{p$_2$}) = I_{\nu}(\text{p$_1$}) e^{-\tau_\nu\text{(p$_1$,p$_2$)}} + 
                                      \int_{\text{p$_1$}}^{\text{p$_2$}} j_\nu(s)e^{-\tau_\nu(s,\text{p$_2$})}ds,
  \end{align}
   where $\tau_\nu$ is the optical depth between two points, p$_1$ and p$_2$, along the ray of sight defined as
  \begin{align}
    \tau_\nu(\text{p$_1$,p$_2$}) = \int_{\text{p$_1$}}^{\text{p$_2$}} \alpha_\nu(s) \text{d}s.
  \end{align}
  \subsection{The gas-phase}\label{sec:gas_phase}
  For the gas, e.g., for a selected transition from lower level $i$ to upper level $j$, the emission ($j_{\nu}$) and absorption ($\alpha_{\nu}$) 
  coefficients have the following form:
  \begin{align}
   j_{ij}(\nu) &= \frac{h\nu} {4\pi} N n_{\text{i}}A_{ij} \Phi_{{ij}}(\nu), \\
   \alpha_{ij}(\nu) &= \frac{h\nu} {4\pi} N (n_{\text{j}}B_{ji}-n_{\text{i}}B_{ij}) \Phi_{ij}(\nu).
  \end{align}
  We follow the ansatz of \cite{Rybicki1991} for the non-overlapping
  multi-level line treatment. For a molecule with N levels and a selected transition between lower $i$ and upper $j$ level, we consider 
  the spontaneous downward transition rates $A_{ij}$, the Einstein coefficients $B_{ij}$ for stimulated transition, 
  and the collision rate $C_{ij}$. The collision rate can be calculated from the number density of the collision partner and the
  downward collision rate coefficient $\gamma_{ij}$, which is the Maxwellian average of the collision cross 
  section $\sigma$ measured in laboratory:
  \begin{align}
   C_{ij} = n_{\text{col}}\gamma_{ij}.
  \end{align}
  All molecule data we use in our software package (e.g., frequencies, transition rates $A_{ij}$, collision rate coefficients)
  are obtained from the \textit{Leiden Atomic and Molecular Database} (LAMDA) \citep{Schoier2005}.
  As a consequence, \textit{Mol3D} uses the same common input format for the molecule data as many other line radiative transfer programs available 
  like RADEX \citep{vanderTak2007} or URAN(IA) \citep{Pavlyuchenkov2004}. This approach offers the advantage of easily 
  extending the code for several common molecules available in the LAMDA database (3 atomic and 33 molecular species at the time of writing).
  The dominant broadening effects are thermal and turbulent
  broadening. Hence, we assume a Gaussian line profile function, but in principle any other profile function can be included as well,
  \begin{align}
    \Phi_{ij}(\nu) &= \frac{c}{\text{v}_{\text{tot}}\nu_{ij}\sqrt{\pi}} \exp \left(-\frac{c^2(\nu - \nu_{ij})^2}{\text{v}_{\text{tot}}^2\nu_{ij}^2}\right),
  \end{align}
  where $\nu_{ij}$ is the central frequency of the transition and c the speed of light. The parameter $\text{v}_{\text{tot}}$ 
  gives the total line broadening factor taking into account the assumed microturbulent velocity $\text{v}_{\text{turb}}$
  and the kinetic velocity $\text{v}_{\text{kin}}$ due to the kinetic temperature $T_{\text{kin}}$:
  \begin{align}
    \text{v}_{\text{tot}} &= \sqrt{\text{v}_{\text{kin}}^2  + \text{v}_{\text{turb}}^2} 
    = \sqrt{ \frac{2kT_{\text{kin}}} {m_{\text{mol}} } + \text{v}_{\text{turb}}^2}\label{eq:broaden}.
  \end{align}
  In our code, the microturbulent velocity $\text{v}_{\text{turb}}$ can either be set to a common value, 
  $\sim$ 0.1 - 0.2 km/s for protoplanetary disks \citep{Pietu2007,Hughes2011},
  or for example in the case of an outcome of (M)HD/MRI simulations, it can be provided as an input for every grid cell.\\ 
  Finally, to describe the full line radiative transfer problem, a
  set of balance equations is needed:
  \begin{align}
    \sum_{i>j}[n_iA_{ij} + (n_iB_{ij} - n_jB_{ji})\bar{J}_{ij}] &- \sum_{i<j}[n_jA_{ji} + (n_jB_{ji} - n_iB_{ij})\bar{J}_{ji}] \label{eq:balance}\\
    &+ \sum_{i} [n_iC_{ij} - n_jC_{ji}] = 0 \nonumber.
  \end{align}
  In fact, Eq. \ref{eq:balance} is not  one single equation, but a whole set of linear equations, one for every energy level. In
  our code this linear matrix equation is solved with a simple Gaussian elimination algorithm. 
  This set has to be solved locally (for every grid cell), but
  it has a global character due to its dependency on the line integrated mean intensity $\bar{J}_{ij}$ defined as
  \begin{align}
    \bar{J}_{ij} = \frac{1}{4\pi} \int d\Omega \int I_{\nu}\Phi(\nu)d\nu. \label{eq:jbar}
  \end{align}
  Equations \ref{eq:balance} and \ref{eq:jbar} are directly coupled to Eq. \ref{eq:RTE}.\\ 
  To solve the line radiative transfer problem, one needs to start with estimated values for the level populations, 
  solve the radiative transfer equation, calculate the 
  mean intensity, and then solve Eq. \ref{eq:balance} again to create a new set of level populations. Then one needs to iterate until the correct 
  level populations have been found. 
\subsubsection{Level populations}\label{sec:levelpop}
  The main problem in radiative line transfer is the calculation of the level populations. Equation (\ref{eq:balance}) shows that the
  level populations are directly coupled to the mean intensity and therefore to the gas temperature, 
  which is unknown for complex structures as in the case of protoplanetary disks.\\
  We implemented dissimilar algorithms to calculate the level populations with three approximate methods, for example, local thermodynamical equilibrium (LTE),
  full escape probability (FEP), and large velocity gradient (LVG). As \citet{Pavlyuchenkov2007} have shown for the case of
  protoplanetary disks, for the most common molecules LVG is a very accurate method of calculating the level populations and gives a 
  very good ratio between computational efficiency and reliability of the results.
\subsubsection*{LTE level populations}
  In this assumption the level populations follow a Boltzmann distribution and the excitation temperature of the 
  selected transition is assumed to be equal to the 
  kinetic temperature $T_{\text{exc}} = T_{\text{kin}} $:
  \begin{align}
    \frac{n_{\text{i}}}{n_{\text{j}}} &= \frac{g_{\text{i}}}{g_{\text{j}}} \exp \left(\frac{h\nu_{{ij}}}{kT_{\text{kin}}}\right).
  \end{align}
  Here $g_{\text{i}}$ and $g_{\text{j}}$ are the statistical weights of the i-j line transition and $\nu_{{ij}}$ the corresponding
  central frequency.
\subsubsection*{LVG level populations}\label{sec:LVG}
  In the case of protoplanetary disks, the radial velocity gradients are usually much larger than the local thermal and microturbulent 
  velocities \citep{Weis2005,Castro-Carrizo2007}. This means that photons emitted at a certain disk region can
  only interact and consequently get absorbed locally.
  In this approach, the mean intensity $J$ is approximated from the local source function $S$ in combination with an external radiation field 
  $J_{\text{ext}}$:
  \begin{align}
   J &= (1-\beta)S +\beta J_{\text{ext}},\\
   0 &< \beta \le 1.
  \end{align}
  The quantity $\beta$ gives the probability of a photon beeing able to escaping the model. 
  We note that in the optically thin case a photon can escape the model without
  interaction. For this case, $\beta$ equals $1$ and the LVG method would be equal to the full escape probability method. 
  Therefore, the FEP method (also included in \textit{Mol3D}) uses the same formalism as the LVG method, but with $\beta$ always set to $1$
  \citep[see also][]{vanderTak2007}.
  To calculate the local $\beta$, our program uses  Eq. \ref{eq.betalvg} introduced by \citet{Mihalas1978,deJong1980} for accretion disks in general,
  \begin{align}
   \beta = \frac{1-\exp(-\tau)}{\tau} \label{eq.betalvg},
  \end{align}
  where $\tau$ is the effective optical depth of the observed line.
  We note that this formula is well suited in the case of proto-circumstellar disks. For other geometries and scenarios
  other approximations might give better results (see, e.g., \citet{deJong1975} or \citet{Osterbrock2006}).
  At this point an approximation of the optical line depth is made. As it is connected to the local velocity field $V$,
  microturbulence $\text{v}_\text{turb}$, and gas density, we use the following expression \citep[see also][]{Pavlyuchenkov2007}:
  \begin{align}
   \tau(\nu) = \alpha(\nu)R\sqrt{\frac{2}{3}\frac{\text{v}_\text{tot}}{V}}.
  \end{align}
\subsection{Benchmarks}
  We demonstrate the applicability of the code \textit{Mol3D}. For this purpose we compare
  its results to those obtained with the line radiative transfer code URAN(IA) \citep{Pavlyuchenkov2004} 
  and the dust continuum radiative transfer code MC3D \citep{Wolf1999,Wolf2003c}.  
  \subsubsection{URAN(IA)}\label{sec:mol3dtourania}
\begin{figure}
  \includegraphics[width=\columnwidth]{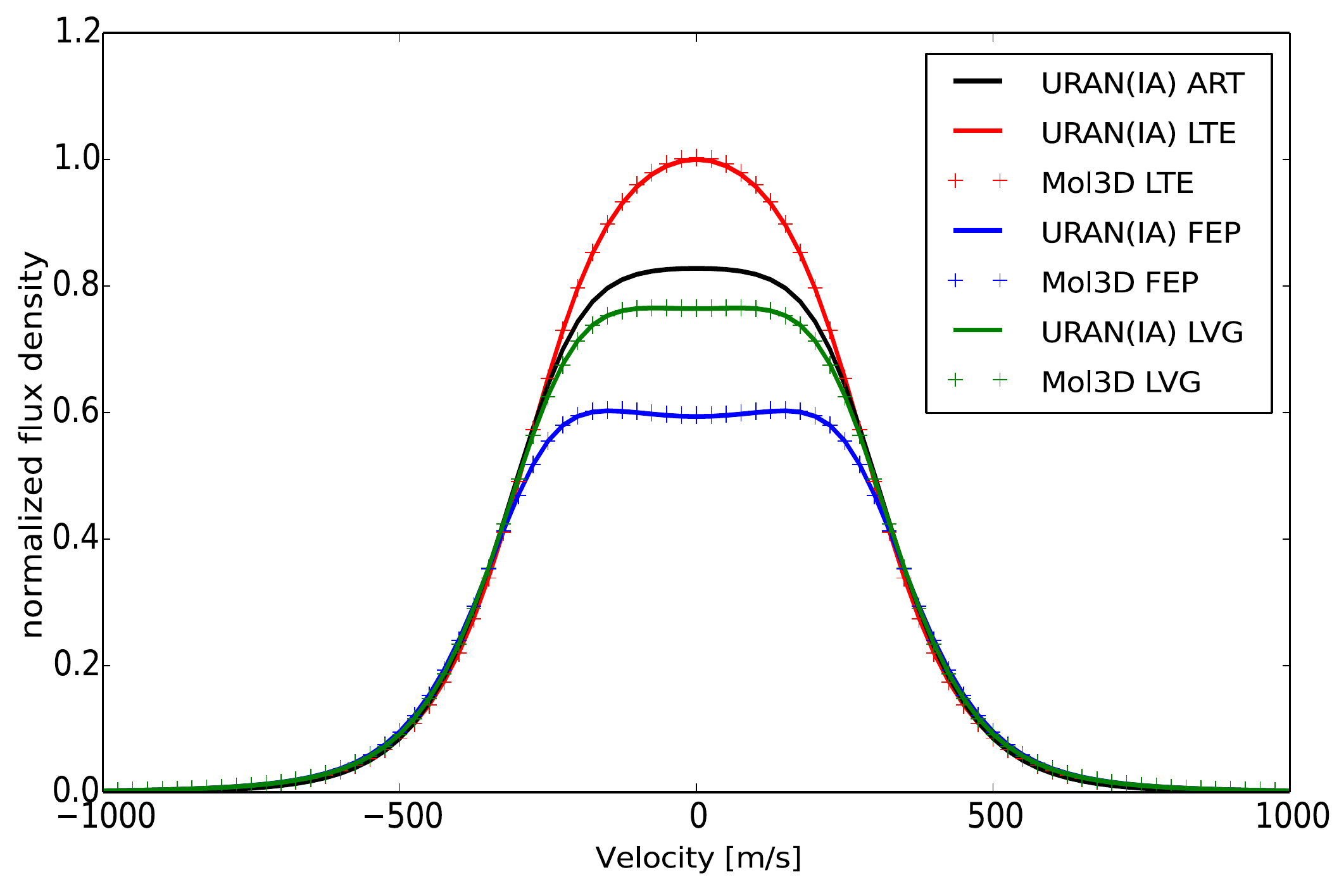}
  \caption{HCO$^{+}$ (4-3) transition of a protoplanetary disk, oriented \textit{\emph{face-on}}. Different colors are obtained with different methods. 
           Solid lines 
           are obtained using the URAN(IA) software package and dotted lines represent solutions obtained with Mol3D. Shown is the normalized 
           flux over velocity. Both codes produce comparable results within acceptable differences (less than 0.5\%). LVG is a good 
           approximation in the case of circumstellar disks \citep{Pavlyuchenkov2007}.
           }\label{pic:urania-mol3d}
  \end{figure}

  URAN(IA) is a 2D radiative transfer code for molecular lines. 
  It features a Monte Carlo
  algorithm for calculating the mean intensity and an accelerated lambda iteration (\textbf{ALI}) method for self-consistent molecular
  excitation calculations. This code has been extensively tested in 1D and 2D, 
  for example against the RATRAN code \citep{Hogerheijde2000}. It successfully passed all tests formulated in the benchmark
  paper for N-LTE molecular radiative transfer by \citet{vanZadelhoff2002}. 
  It has been used in several applications, for example modeling the starless core L1544 \citep{Pavlyuchenkov2004},
  and in dedicated parameter space studies of molecular line formation of molecular line formation of prestellar cores \citep{Pavlyuchenkov2008}.\\
  In this code, several approximate methods of calculating the level populations are included. In a comprehensive study, ~\citet{Pavlyuchenkov2007}
  have shown the reliability of these methods in protoplanetary disk scenarios.\\
  In Sect. \ref{sec:sequence_mol3d} we introduced the different methods of  calculating level populations implemented in \textit{Mol3D}.
  In Fig. \ref{pic:urania-mol3d} we present the spectrum of the HCO$^{+}$ (4-3) transition of a typical protoplanetary disk 
  (M$_{\rm disk}$ = 0.07~M$_{\odot}$), comparing the different methods assuming a uniform N(HCO$^{+}$)/N(H) ratio of 1$\cdot$10$^{-8}$. 
  For simplicity, we choose a flared disk model around a T~Tauri star utilizing a temperature distribution with a gradient in radial and vertical direction
  and a power-law model for the density distribution adapted from \citet{Shakura1973}: 
  \begin{align}
    \rho_{dust}(r,z) &\sim \left(\frac{r}{100 \text{AU}}\right)^{-\alpha} \exp\left(\frac{z^2}{h(r)^2}\right), \label{eq:shakura}\\
    h(r) = 10&\text{AU}\cdot \left(\frac{r}{100 \text{AU}}\right)^{\beta}.\nonumber
  \end{align}
  The disk is orientated \textit{\emph{face-on}}. We assume pure Keplerian rotation and neglect dust re-emission.
  All spectra are normalized to the maximum value of the LTE solution. To allow comparison,
  the resulting spectra obtained with the URAN(IA) code are also included. \\
  URAN(IA) and \textit{Mol3D} produce comparable HCO$^{+}$ (4-3) spectra with differences of less than about 0.5\%. 
  The most realistic solution (black line) is obtained using the accelerated Monte Carlo (\textbf{ART}), 
  to classify the approximation methods.
  As discussed in greater detail in  Appendix \ref{sec:levelpop}, the LTE method overestimates the net flux and the FEP method underestimates it. 
  The fluxes in the line obtained with the LVG method are in between those obtained with the other methods
  and are closely  comparable to the ART method. This result has also been found by \citet{Pavlyuchenkov2007} and we refer the interested reader to
  their study.  
 \subsubsection{MC3D}\label{sec:mol3dtomc3d}
  MC3D is a 3D continuum radiative transfer code that uses a Monte Carlo method to calculate self-consistent dust temperature distributions,
  continuum re-emission/scattering maps, and SEDs.
  Among a wide variety of radiative transfer studies it has been applied extensively for the analysis of multiwavelength 
  high-angular-resolution observations of circumstellar disks \citep[e.g.,][]{Schegerer2008,Sauter2011,Madlener2012,Grafe2013a}. 
  MC3D was successfully tested against other continuum RT codes \citep[e.g.,][]{Pascucci2004}.\\
  In this section, we compare \textit{Mol3D} with MC3D for a typical protoplanetary disk scenario.
  The density distribution is described by Eq. \ref{eq:shakura}, assuming a total disk mass of 0.04~M$_{\odot}$. The
  temperature structure is calculated self-consistently with each code using their Monte Carlo scheme. 
  Both codes use a similar approach based on the assumption of local thermal equilibrium and immediate temperature correction 
  as proposed by \citet{Bjorkman2001}.
  \begin{figure}
  \includegraphics[width=\columnwidth]{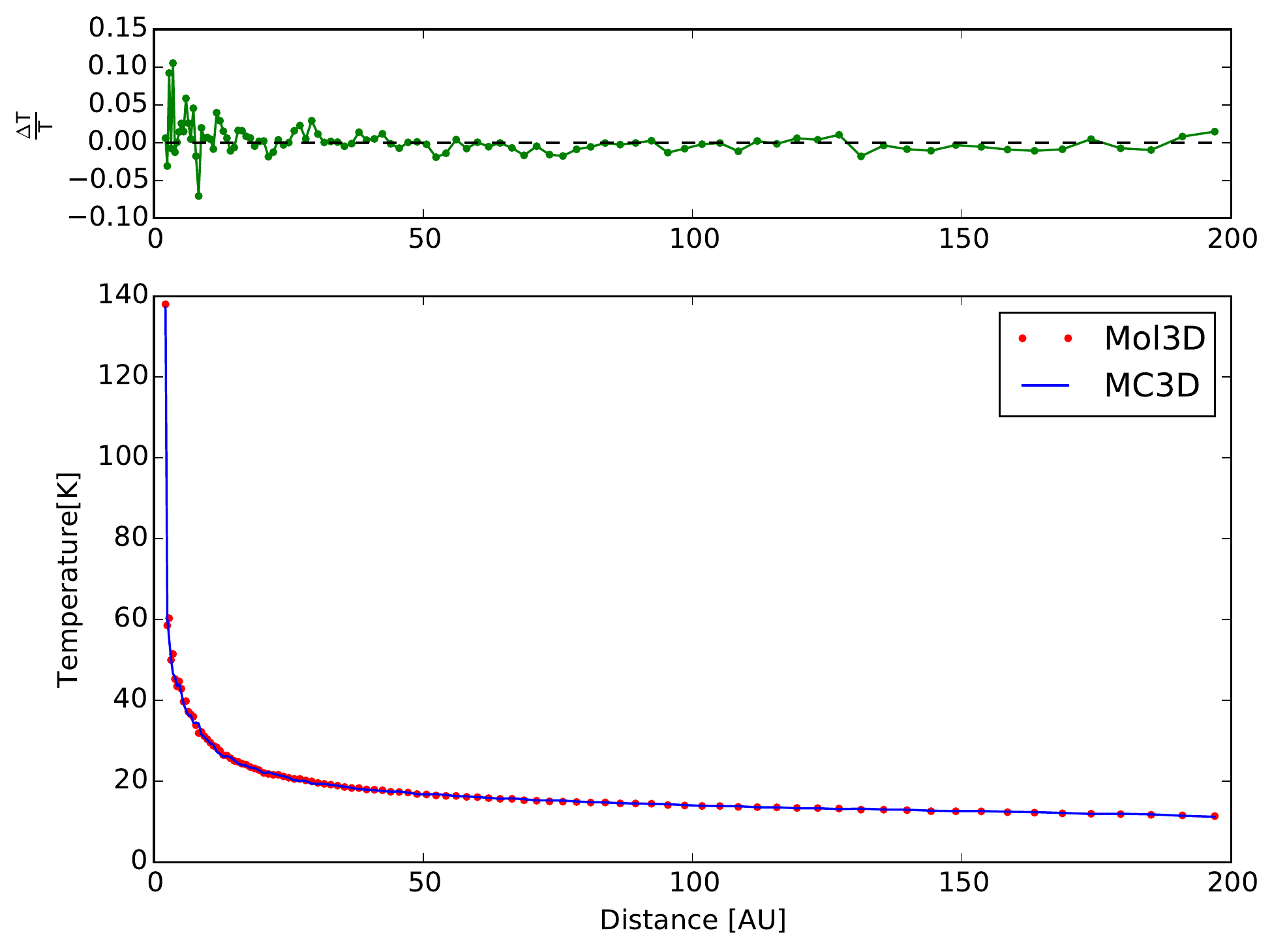}
        \caption{Midplane temperature of a typical protoplanetary disk around a pre-main-sequence T~Tauri star. This disk inner radius amounts
                 to 2~AU and the outer radius to 200~AU, respectively. The total disk mass amounts to 0.04~M$_{\odot}$.
                 Mol3D (red) and MC3D (blue) obtain  comparable results. 
                 In this case, the differences depend significantly on the optical disk properties and subsequently on the number
                 of photon packages simulated in the Monte Carlo process. Thus, the error is higher in the dense region at the inner rim of the
                 disk (max. 10\%) and lower at the outer disk regions (max. 2\%).
                 }\label{pic:mc3d-mol3d-temp}
  \end{figure}\\
  The resulting midplane temperature around a 
  pre-main-sequence T~Tauri star is shown in Fig.~\ref{pic:mc3d-mol3d-temp}. 
  The calculation of the temperature in the dense midplane is  one of the most crucial RT problems, 
  because these disk regions can hardly be reached by stellar photons directly. 
  Thus, it is mostly heated indirectly by thermal re-emission radiation from the innermost disk regions and upper disk layers.
  Both codes produce very smooth and similar temperature distributions. The statistical nature of the Monte Carlo method
  results in deviations of about 10\% for the innermost disk parts 
  and less than 2\% for the outer regions, which is mainly due to the probability that the photon packages can reach these regions.\\
  We also compared the dust re-emission/scattered light maps and SEDs and find that both codes produce  comparable 
  results with maximum deviations of about 10\%.
\end{document}